\documentclass[letterpaper]{article}
\usepackage{aaai2026}
\usepackage{times}
\usepackage{helvet}
\usepackage{courier}
\usepackage[hyphens]{url}
\usepackage{graphicx}
\usepackage{natbib}
\usepackage{caption}
\usepackage{newfloat}
\usepackage{listings}
\usepackage[compress]{cite}
\usepackage{amsmath,amssymb,amsfonts} % 支持数学公式
\usepackage{subcaption}
\usepackage{graphicx} % 引入graphicx包来插入图片
\usepackage{textcomp}
\usepackage[ruled,vlined,linesnumbered]{algorithm2e}
\usepackage{xspace}
\usepackage{booktabs}
\usepackage{multirow}
\usepackage{subcaption}
\usepackage[table]{xcolor} % 表格颜色（如果需要）
\usepackage{tcolorbox}
\usepackage{arydshln} % 支持虚线 \hdashline
\usepackage{bm}
\newcommand{\G}{\mathbin{\mathbf{G}}}
\newcommand{\F}{\mathbin{\mathbf{F}}}

\newcommand{\oomit}[1]{}

{\bfseries}{\rmfamily} 
{\bfseries}{\rmfamily} 
{\bfseries}{\rmfamily}
{\bfseries}{\rmfamily}
\newtheorem{mydefinition}{Definition}{\bfseries}{\rmfamily}
{\bfseries}{\rmfamily}

\allowdisplaybreaks

\title{RESTL: Reinforcement Learning Guided by Multi-Aspect Rewards for Signal Temporal Logic Transformation}
\author{ 
  \fontfamily{ptm}\selectfont  % 设置为Times字体（可选：ppl=Palatino, bch=Chancery等）
  Yue Fang\textsuperscript{\rm 1,2}, Zhi Jin\textsuperscript{\rm 1,2}\thanks{Corresponding author}, Jie An\textsuperscript{\rm 3,4}\footnotemark[1], {\bf Hongshen Chen\textsuperscript{\rm 5}}, {\bf Xiaohong Chen\textsuperscript{\rm 6}}, {\bf Naijun Zhan\textsuperscript{\rm 1,2}} \\ 
  \small
  \textsuperscript{1}School of Computer Science, Peking University, Beijing, China \\ 
  \textsuperscript{2}Key Laboratory of High Confidence Software Technologies (PKU), MOE, China \\ 
  \textsuperscript{3}National Key Laboratory of Space Integrated Information System \\ 
  \textsuperscript{4}Institute of Software, Chinese Academy of Sciences, Beijing, China \\ 
  \textsuperscript{5}JD.com, Beijing, China \\
  \textsuperscript{6}East China Normal University, Shanghai, China \\ 
  \vspace{0.5em}
  y.fang@stu.pku.edu.cn, zhijin@pku.edu.cn, anjie@iscas.ac.cn
}

\begin{document}
\maketitle

\begin{abstract}
Signal Temporal Logic (STL) is a powerful formal language for specifying real-time specifications of Cyber-Physical Systems (CPS).  %and 
%Recently, t
Transforming %such 
specifications written in natural language into STL formulas automatically has %gained
attracted increasing attention. Existing rule-based methods depend heavily on rigid pattern matching and domain-specific knowledge, limiting their generalizability and scalability.
% Recent approaches using large language models (LLMs) with supervised fine-tuning (SFT) often generate STL formulas that are inconsistent with the input natural language. 
%More 
Recently,
% supervised fine-tuning (SFT) of large language models (LLMs) has been applied to the natural language to STL transformation task. 
Supervised Fine-Tuning (SFT) of large language models (LLMs) has been successfully applied to transform natural language into STL. 
% However, formulas generated by these SFT-based methods often exhibit inconsistencies with the intended meaning of the input. This issue fundamentally arises because the single-objective training goal of SFT lacks mechanisms to impose fine-grained constraints across multiple dimensions, such as semantic fidelity, atomic proposition alignment, and logical structure.
However, the lack of fine-grained supervision on 
% semantic fidelity, atomic proposition alignment, logical structure, etc
semantic fidelity, atomic proposition correctness, and formula readability often leads SFT-based methods to produce formulas misaligned with the intended meaning.
To address these issues, we propose RESTL, a reinforcement learning (RL)-based framework for the transformation from natural language to STL. RESTL introduces multiple independently trained reward models that provide fine-grained, multi-faceted feedback from four perspectives, i.e., atomic proposition consistency, semantic alignment, formula succinctness, and symbol matching.
These reward models are trained with a curriculum learning strategy to improve their feedback accuracy, and their outputs are aggregated into a unified signal that guides the optimization of the STL generator via Proximal Policy Optimization (PPO).
% These reward signals are first trained through a curriculum learning strategy, guiding each reward model progressively from simple to complex samples to learn evaluation criteria in different dimensions. Subsequently, we integrate these rewards into a composite optimization objective that directs the Proximal Policy Optimization (PPO) algorithm to fine-tune the STL generation model via reinforcement learning.
Experimental results demonstrate that RESTL significantly outperforms state-of-the-art methods in both automatic metrics and human evaluations.
% The code is available in the supplementary material.

\end{abstract}

\section{Introduction} \label{sec:introduction}

% Signal Temporal Logic (STL) \cite{STL} is an extension of classical Linear Temporal Logic (LTL) \cite{LTL} by allowing to specify properties of dense-time real-valued signals. 
% Signal Temporal Logic (STL)，an extension of classical Linear Temporal Logic, is currently a well-known specification language with powerful semantics for formally specifying requirements with time constraints for Cyber-Physical Systems (CPS) by allowing to specify properties of dense-time real-valued signals, as it can enhance the engineering requirements process by ensuring greater precision and rigor~\cite{BartocciMNN22}.
Signal Temporal Logic (STL)~\cite{maler2004monitoring}, an extension of classical Temporal Logic (TL)~\cite{pnueli1977temporal}, is currently a well-known specification language for formally specifying requirements of cyber-physical systems (CPS) with dense-time real-valued signals.
%\cite{BartocciMNN22}.
% , thereby enhancing the engineering requirements process by ensuring greater precision and rigor~\cite{BartocciMNN22}.
STL has been applied to critical tasks such as model checking and runtime monitoring of CPS in both academia and industry~\cite{maierhofer2020formalization,tellex2020robots}. 
%These applications have significantly expanded its practical relevance, leading to increasing attention from both academia and industry~\cite{maierhofer2020formalization,tellex2020robots}.
% As a result, STL now plays an important role in the design and verification of CPS.
However, most of the requirements regarding the timing constraints of CPS are typically specified informally in domain documentation written in natural language by domain experts. 
The lack of an effective method for transforming informal requirements into corresponding formal STL specifications has become a critical challenge, limiting its broader adoption in real-world CPS design and analysis.

% This becomes a major bottleneck to the wider acceptance of STL, this formal technique, in the actual design of CPS, as it is not easy to transform the informal requirements in natural language into formal formulas \cite{LearningLTL19}.
%控制生成等形式化方法是需要STL的，但目前这类的specification是以自然语言形式表达，如何把自然语言转化成STL公式的自动化过程是一个瓶颈。STL既是规约语言，后续的是重要的，比如model checking。
%如何将非形式化的需求转换成形式化的STL是目前的主要瓶颈，

% 已有的方法，不要用传统。
% 要表达一个好的STL的公式，要从多个地方来捕捉多个维度的信息，所以它很依赖于自然语言理解的能力
% 各维度信息的提取。
% 换一下形式就不起作用。LLM的自然语言理解很厉害，但它不是STL的专家。
\begin{table}[!t]
\centering
\renewcommand{\arraystretch}{1.4}
\small
\resizebox{0.95\linewidth}{!}{
\begin{tabular}{p{8.3cm}}
\toprule
\textbf{Case 1 – Temporal Semantic Error} \\
\hline
\textit{Natural Language Description:} \\
During 10–150 time units, if signal $z_1$ is less than 0.2, then signal $z_2$ remains less than 0.3 from 1 to 3 time units later. \\
\hdashline
\textit{LLaMA 3-8B (fine-tuned):} \\[-1pt]
$\G_{[10,150]}((z_1 < 0.2) \rightarrow \underline{\F}_{[1,3]}(z_2 < 0.3))$ \\
\hdashline
\textit{Ground Truth:} \\[-1pt]
$\G_{[10,150]}((z_1 < 0.2) \rightarrow \underline{\G}_{[1,3]}(z_2 < 0.3))$ \\[2pt]
\toprule
\textbf{Case 2 – Atomic Proposition Error} \\
\hline
\textit{Natural Language Description:} \\
% If vehicle speed $v$ stays within 60–80 km/h for at least 10 consecutive seconds within the next minute, then the vehicle must not exceed 100 km/h in the next two minutes. \\
If the rear radar detects an obstacle and the reverse gear is engaged, then the rear brake signal \texttt{brake\_rear} should be activated within 2 seconds. \\
% If the rear radar detects an obstacle and the reverse gear is engaged, then the rear brake signal brake\_rear should be activated within 2 seconds. \\
\hdashline
\textit{LLaMA 3-8B (fine-tuned):} \\[-1pt]
% $\F_{[0,60]}(\G_{[0,10]}(60 \leq v \leq 80)) \rightarrow \G_{\textcolor[RGB]{220,20,60}{[60,420]}}(v \leq 100)$ \\
% $\G_{[0,T]}(\uwave{\texttt{radar\_rear.detect\_obstacle}} \rightarrow $
% \\[-2pt]$\F_{[0,2]}(\texttt{brake\_rear} = 1))$ \\
$\G_{[0,T]}(\underline{\texttt{radar\_rear.detect\_obstacle}} \rightarrow $
\\[-2pt]$\F_{[0,2]}(\texttt{brake\_rear} = 1))$ \\
\hdashline
\textit{Ground Truth:} \\[-1pt]
% $\F_{[0,60]}\big(\G_{[0,10]}(60 \leq v \leq 80) \rightarrow \G_{\textcolor[RGB]{46,139,87}{[0,120]}}(v \leq 100)\big)$ \\
$\G_{[0,T]}((\underline{\texttt{radar\_rear.detect\_obstacle}} \land \underline{\texttt{gear\_rev} = 1})$ $ \rightarrow  \F_{[0,2]}(\texttt{brake\_rear} = 1))$ \\[2pt]
\toprule
\textbf{Case 3 – Formula Redundancy} \\
\hline
\textit{Natural Language Description:} \\
The temperature $T$ is consistently above 22°C during the first 2 hours and then rises above 30°C sometime between 2 and 4 hours. \\
\hdashline
\textit{LLaMA 3-8B (fine-tuned):} \\
% $0 \leq t \leq 120 \rightarrow \G_{[0,120]}(T[t] > 22) \land \textcolor[RGB]{220,20,60}{\F_{[120,240]}(T[t] > 30)}$ \\
$\underline{(\G_{[0,120]}(T > 22) \land \F_{[0,240]}(T > 30) \land \G_{[120,240]}(T > 30))}$ \\[2pt]
\hdashline
\textit{Ground Truth:} \\
$\underline{\G_{[0,2]}(T > 22) \land \F_{[2,4]}(T > 30)}$ \\
\bottomrule
\end{tabular}
}
% \caption{Examples of common errors in NL-to-STL transformation by the fine-tuned LLM. \textcolor[RGB]{46,139,87}{Green} text in ground truth indicates the correct STL parts. \textcolor[RGB]{220,20,60}{Red} text in LLaMA3 indicates inconsistencies corresponding to the cases.}
\caption{Examples of common errors in NL-to-STL transformation by the fine-tuned LLM. Underlined parts of formulas indicate incorrect outputs in LLaMA3 and correct ones in the ground truth.}
\label{tab:introduction}
\vspace{-0.4cm}
\end{table}

Manually writing accurate STL formulas is a burdensome task for domain experts because it is time-consuming and error-prone. 
Consequently, many studies have explored automatic methods for transforming natural language descriptions into STL specifications to alleviate this burden and improve accuracy. 
% A common line of work adopts rule-based or pattern-based transformation methods, which rely heavily on meticulously predefined pattern templates and require inputs strictly structured in natural language to match these patterns exactly. 
Among existing methods, rule-based and pattern-based approaches are widely adopted. 
% For example, ~\cite{LignosRFMK15, ghosh2016arsenal} transform natural language sentences into intermediate forms using fixed patterns. 
% For example, Lignos et al.\cite{LignosRFMK15} and Ghosh et al.\cite{ghosh2016arsenal} transform natural language sentences into intermediate forms using fixed patterns.
For example, fixed patterns have been used to transform natural language into intermediate forms in previous work~\cite{LignosRFMK15, ghosh2016arsenal}.
Then, with a set of manually designed rules, the intermediate forms are further transformed into temporal logic formulas.
% Such methods require substantial domain-specific expertise and considerable manual intervention, severely restricting their adaptability and making them inadequate for handling diverse, ambiguous, or complex linguistic scenarios commonly encountered in real-world applications.
These methods rely on meticulously crafted templates, which require substantial expert effort and steep learning curves~\cite{kulkarni2013new}. Moreover, they are typically limited to highly restrictive and structured natural language expressions that strictly match predefined patterns.

Advancements in natural language processing (NLP), particularly the impressive capabilities demonstrated by large language models (LLMs), have sparked a strong interest in employing these %technologies 
techniques 
for transforming natural language into STL. 
Recent techniques have explored different strategies for tackling the Natural Language to STL (NL-to-STL) transformation problem.
% DeepSTL~\cite{HeBNIG22} relies on grammar-guided data synthesis and trains transformer-based models to learn the mapping from natural language to STL.
For example, DeepSTL~\cite{HeBNIG22} uses grammar-guided data synthesis to train transformer-based models that learn to map natural language inputs to STL formulas.
% NL2TL\cite{ChenGZF23} similarly constructs synthetic NL–STL pairs and performs instruction tuning on large language models, enabling them to generate STL formulas under natural language guidance.
NL2TL~\cite{ChenGZF23} is a method based on Supervised Fine-Tuning (SFT) that constructs synthetic NL–STL pairs and performs instruction tuning on LLMs, relying solely on paired data as supervision to allow the generation of STL formulas from natural language instructions.
KGST~\cite{fang2025enhancingtransformationnaturallanguage} is a two-stage method that first fine-tunes an LLM on NL–STL pairs and then refines its outputs using external knowledge.
% Due to the coarse-grained supervision and fixed training objectives, these methods offer limited guidance on the fine-grained semantic and structural decisions required for accurate STL generation.

% However, because of their coarse-grained supervision and static training objectives, these methods provide limited guidance for accurate STL generation.
% While these methods have shown promise
While current methods have made notable progress in automating the transformation from natural language to STL, their accuracy in generating correct STL formulas remains insufficient. 
Table~\ref{tab:introduction} illustrates representative examples of common errors made by a fine-tuned LLaMA 3-8B model compared to the ground-truth. Specifically, 
Case 1 shows a temporal semantic error where the model fails to capture the meaning of ``remains'', leading to the incorrect temporal operator ``$\F$''. 
Case 2 reflects atomic proposition misalignment, as the model omits the condition ``reverse gear is engaged''.
Case 3 illustrates formula redundancy, as the model generates unnecessary temporal constraints.
% Case 1 demonstrates a temporal semantic error, Case 2 highlights a misalignment of atomic propositions, and Case 3 illustrates redundancy in the generated formula. 
These issues reflect the limitations of existing SFT-based approaches, which often rely on coarse-grained supervision and static training objectives, providing limited guidance for learning the precise semantic information required for accurate STL generation.
To address the limitations arising from coarse-grained supervision, we propose RESTL, a multi-aspect reward-guided reinforcement learning framework for transforming natural 
language into STL.
%Signal Temporal Logic (STL). 
RESTL introduces four complementary reward metrics to provide fine-grained supervision from multiple perspectives: (1) Atomic Proposition Alignment, which checks whether all key variables are accurately captured; (2) Templated Natural Language Similarity, which evaluates semantic alignment by focusing on semantic content and logical consistency; (3) Formula %Conciseness
Succinctness, which measures
the difference between the length of the output formula and 
%outputs that are closer in length 
 its ground truth to improve succinct yet faithful expressions. 
 % Normally, shorter the difference is better; 
Normally, a smaller difference is better.
 and (4) STL-level Similarity, which measures the similarity between the generated and reference formulas, providing global supervision for the formula generation. 
Each metric corresponds to a lightweight reward model, which is trained via preference learning on multiple generated STL outputs to provide evaluative feedback for guiding the generator.
% These metrics are instantiated as lightweight reward models trained via preference-based learning on paired STL outputs. 
To improve learning accuracy, we employ a curriculum learning strategy that gradually increases task difficulty for each reward model. 
% The outputs of these models are then aggregated into a unified scalar signal, which is used to optimize the STL generator with Proximal Policy Optimization (PPO), balancing reward maximization and training stability through KL-regularized policy updates.
Finally, the outputs of these reward models are aggregated into a unified scalar signal to optimize the STL generator using Proximal Policy Optimization (PPO)~\cite{schulman2017proximal}. During this optimization, we incorporate a Kullback-Leibler (KL) regularization strategy to constrain drastic policy updates while maximizing the reward and enhancing training stability.

Experimental results show that our RESTL framework significantly outperforms baseline methods in both automatic and human evaluations. It achieves higher accuracy in transforming natural language descriptions into STL 
%Signal Temporal Logic (STL) 
formulas, with better semantic alignment and readability. 
% These improvements confirm the effectiveness of multi-aspect reward modeling combined with curriculum learning in guiding the model to generate more precise STL specifications. 
%representations.

%In general, 
In summary, our main contributions include: %can be summarized as follows.
\begin{itemize}
\item We propose RESTL, the first reinforcement learning framework for transforming natural language into STL. RESTL learns from multiple dimensions of feedback, including atomic proposition consistency, semantic alignment,
formula succinctness, and symbol matching.
% \item We introduce a difficulty-aware curriculum learning strategy based on reward-driven metrics, allowing the STL generator to progressively learn from easier to harder cases. This curriculum improves training stability and model generalization on complex STL constructions.
\item We introduce a curriculum learning strategy to train each reward model from easier to harder examples, improving the accuracy of the reward and training stability.
\item Experimental results on two datasets show that RESTL outperforms baselines in both automatic and human evaluations.
\end{itemize}
% 总结一下contribution

% \section{Signal Temporal Logic formulas}

\section{Related Work}
% Related work contains two parts. The first part reviews literature on the temporal logic transformation. The second part discusses the methods that can be used to enhance the performance of LLMs in certain tasks and the application of LLMs in generating formulas.
% \subsection{From Natural Language to TL and STL} 
% Many researchers have made efforts to transform natural language (NL) into Temporal Logic (TL) specifications. 
% \myparagraph{From NL to TL} 
% 
In this section, we present the most relevant related work, more details can be found in the Appendix A.
Many efforts have been made to transform natural language (NL) into TL specifications. 
For example, a catalog of temporal logic formulas that capture common specification patterns in the design of concurrent and reactive systems was proposed in~\cite{dwyer1999patterns}.
% For example, \cite{dwyer1999patterns} proposes a catalog of TL formulas that capture common specification patterns in the design of concurrent and reactive systems.
Controlled English has also been transformed into TL through the use of syntactic and grammatical dependency parsing, together with predefined mapping rules~\cite{vzilka2010temporal, santos2018formal}.
% \v{Z}ilka et al.\cite{vzilka2010temporal} and Santos et al.\cite{santos2018formal} transformed controlled English into LTL by applying syntactic and grammatical dependency parsing combined with predefined mapping rules.
% \cite{vzilka2010temporal} and \cite{santos2018formal} transform controlled English into LTL using syntactic and grammatical dependency parsing together with predefined mapping rules.
% Similarly, the ARSENAL framework~\cite{ghosh2016arsenal} uses NLP techniques such as n-gram analysis and dependency parsing to transform natural language requirements into formal specifications, including TL.
Although these methods are effective in specific domains, they rely on handcrafted rules and restricted language inputs, which limit their ability to handle more diverse and complex expressions. 
To overcome these limitations, the nl2spec method~\cite{CoslerHMST23} integrates human feedback and LLMs to automatically derive TL formulas.
% To address these limitations, \cite{CoslerHMST23} proposes nl2spec, which combines human feedback and large language models (LLMs) to derive LTL formulas.
However, these TL-focused methods are not readily extended to STL, because STL introduces real-valued signals and continuous-time constraints, which pose challenges not typically addressed by traditional TL.
% which are beyond the expressive power of traditional TL.

% \myparagraph{From NL to STL} 
As an extension of TL that incorporates real-valued dense-time signals, STL has gained widespread use in both academia and industry to meet the requirements of CPS~\cite{MadsenVSVDWDB18}. 
Consequently, numerous efforts have been made to transform natural language into STL. 
For example, DeepSTL~\cite{HeBNIG22} trains a Transformer model using grammar-based synthetic data. 
Although this approach ensures formal consistency, it heavily relies on handcrafted rules and artificial data, which fail to capture the diversity in real-world natural language. 
% As a result, its generalization capability is limited when applied to open-domain inputs. 
% Dialogue-based approaches such as DialogueSTL~\cite{mohammadinejad2024systematic} improve transformation accuracy through user interaction and transformer models, but still rely on human feedback.
In addition, NL2TL~\cite{ChenGZF23} mitigates the reliance on rule design by fine-tuning a T5 model on NL–TL pairs generated by LLMs. 
KGST~\cite{fang2025enhancingtransformationnaturallanguage} uses a generate-then-refine approach by first fine-tuning LLMs to generate initial STL formulas, then refining them with external knowledge.
However, as supervised fine-tuning methods, both NL2TL and KGST optimize fixed training objectives, lacking fine-grained feedback and struggling to accurately represent the semantics of the input natural language.
% In contrast, reinforcement learning offers the flexibility to dynamically adjust the generation strategy based on targeted feedback, making it more effective for improving NL-to-STL translation quality. 
% Both of these methods rely on static training data and coarse-grained supervision, making them insufficient for capturing the dual challenges of semantic precision and structural expressiveness in STL generation. 
To address these limitations, we propose RESTL, a reinforcement learning-based framework for NL-to-STL transformation that incorporates multi-aspect supervision and curriculum-guided reward modeling to enhance both the accuracy and readability of generated STL formulas.

%？In summary, these methods demand substantial manpower and time to develop the necessary rules, 
%often resulting in inconsistencies and errors in the translation outcomes.
%These methods face issues like inconsistency, limited generalization, and the need for manual intervention. 

% To address the challenges associated with understanding complex nested logic, we construct a new domain-specific dataset with different complexity levels to help explore more efficient solutions.

\section{Preliminary} \label{sec:preliminary}
STL is a widely used formalism for specifying the real-time properties of CPS, such as autonomous vehicles, robotic systems, etc~\cite{maierhofer2020formalization,tellex2020robots}.

Let \( \mathbb{R} \) denote the set of real numbers, and let \( \mathbb{R}_{\geq 0} \) and \( \mathbb{R}_+ \) denote the sets of non-negative and positive real numbers, respectively. We denote \( \mathbb{N}_{\geq 0} \) and \( \mathbb{N}_+ \) the set of non-negative integers and the set of positive integers, respectively.

Given a time horizon \( T \in \mathbb{R}_+ \) and a signal dimension \( d \in \mathbb{N}_+ \), a \emph{$d$-dimensional signal} is a function \( \mathbf{v} : [0, T] \to \mathbb{R}^d \). For any time \( t \in [0, T] \), \( \mathbf{v}(t) \in \mathbb{R}^d \) represents the values of $d$ signal variables at time $t$. Each component may correspond to physical quantities such as \textsf{velocity}, \textsf{RPM}, or \textsf{acceleration}. In this paper, we fix a set $X$ of such variables and refer to one-dimensional signals as signal variables.

\begin{mydefinition}[STL Syntax]
STL formulas \( \varphi \) are constructed from atomic propositions \( \alpha \) as follows:
\begin{align*}
  &\alpha ::= f(x_1, \dots, x_K) > 0 \\
  &\varphi ::= \alpha \mid \bot \mid \neg \varphi \mid \varphi_1 \wedge \varphi_2 \mid \G_I \varphi \mid \F_I \varphi \mid \varphi_1\,\mathcal{U}_I\,\varphi_2
\end{align*}
Here, $\alpha$ represents atomic proposition, where $f$ is a real-valued function over variables $x_1, \dots, x_K \in X$. $I = [l, u] \subseteq \mathbb{R}_{\geq 0}$ is a closed interval with $l < u$, and $l,u\in\mathbb{N}_{\geq 0}$. The temporal operators $\G$, $\F$, and $\mathcal{U}$ denote ``always'', ``eventually'', and ``until'' respectively. 
%Notably, $\F_I \varphi$ and $\G_I \varphi$ can be expressed via $\mathcal{U}$: $\F_I \varphi \equiv \top\,\mathcal{U}_I\,\varphi$ and $\G_I \varphi \equiv \neg\F_I\neg\varphi$. Additional connectives such as ``disjunction'' $\lor$ and ``imply'' $\to$ are defined as syntactic sugar.
\end{mydefinition}

The Boolean semantics of an STL formula are evaluated over a signal $\mathbf{v}$ at time $t$ as follows:
\begin{align*}
   & (\mathbf{v},t) \models \alpha  &  \Leftrightarrow \quad & f(\mathbf{v}(t)) \geq 0 \\
   & (\mathbf{v},t) \models \neg\varphi & \Leftrightarrow  \quad & (\mathbf{v},t) \not\models \varphi \\
   & (\mathbf{v},t) \models \varphi_1 \wedge \varphi_2 & \Leftrightarrow  \quad & (\mathbf{v},t) \models \varphi_1 \wedge (\mathbf{v},t) \models \varphi_2 \\
   & (\mathbf{v},t) \models \G_{[l,u]} \varphi & \Leftrightarrow  \quad & \forall t' \in [t+l,t+u].\, (\mathbf{v},t') \models \varphi \\
   & (\mathbf{v},t) \models \F_{[l,u]} \varphi & \Leftrightarrow  \quad & \exists t' \in [t+l,t+u].\, (\mathbf{v},t') \models \varphi \\
   & (\mathbf{v},t) \models \varphi_1\,\mathcal{U}_{[l,u]}\,\varphi_2 & \Leftrightarrow  \quad & \exists t' \in [t+l,t+u].\, (\mathbf{v},t') \models \varphi_2 \\
   & & & \wedge\, \forall t'' \in [t, t'].\, (\mathbf{v},t'') \models \varphi_1
\end{align*}

% As a representative example, the Automatic Transmission (AT) benchmark---commonly adopted in CPS falsification competitions~\cite{ARCHCOMP22Falsification}---models a vehicle's transmission controller that continuously outputs signals such as \textsf{temperature}, \textsf{velocity}, \textsf{RPM}, etc. 
For example, considering one of safety requirements of a vehicle's transmission controller~\cite{ARCHCOMP22Falsification}: ``\emph{During the next 21 time units, whenever the velocity exceeds 40, the RPM must drop below 2500 within four time units.}'', it can be expressed as an STL formula over real-valued signals as follows:
$
\G_{[0,21]}(\textsf{velocity} > 40 \to \F_{[1,4]}(\textsf{RPM} < 2500)).
$
\section{Method} \label{sec:method}

\begin{figure}[!t]
    \centering
    \includegraphics[width=1\linewidth]{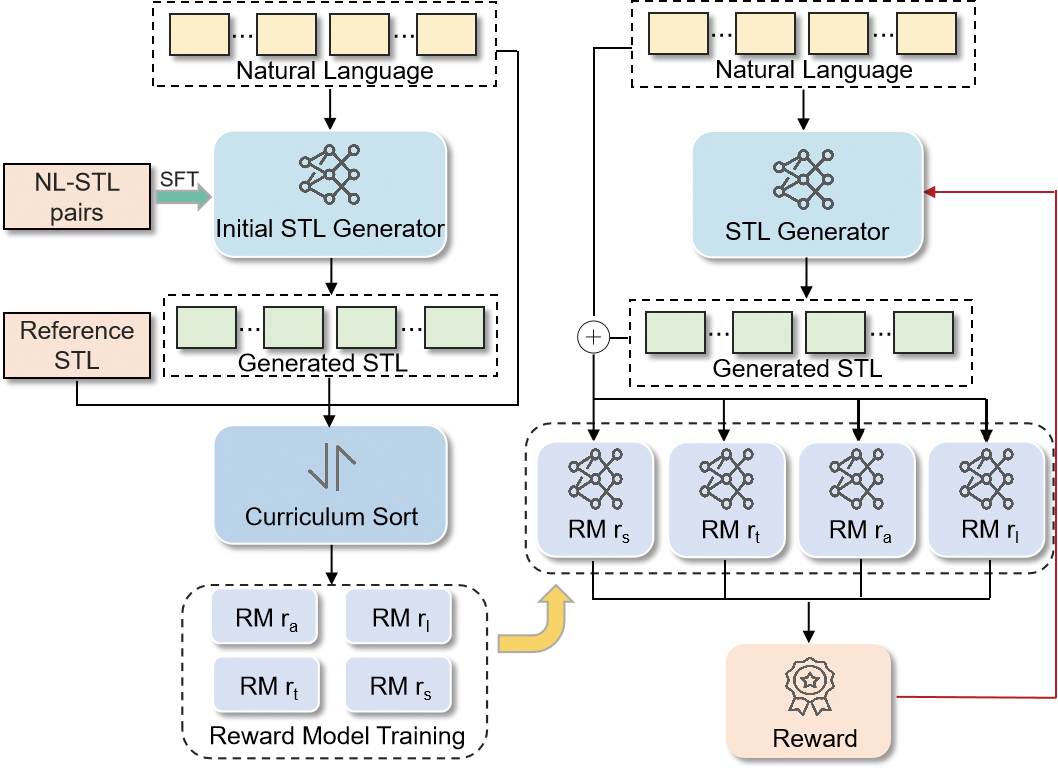}
    \caption{The RESTL framework. First, initialize the STL generator with NL–STL pairs. Then, natural language inputs with both reference and generated STL formulas are used to train multi-aspect reward models based on curriculum learning. Finally, the STL generator is optimized using PPO.}
    \label{fig:framework}
    %\vspace{-0.5cm}
\end{figure}

% \textcolor{red}{In this section, we introduce RESTL, which is a reinforcement learning framework integrating four reward models ..., ......  }
% 介绍一下它是什么，大致是怎么做的
% First 一个基础的模型
In this section, we present RESTL, a reinforcement learning framework for transforming natural language into STL. RESTL integrates four distinct reward models trained through curriculum learning to provide multi-aspect feedback, which is then aggregated to guide the optimization of the STL generator using the PPO algorithm, as illustrated in Figure~\ref{fig:framework}.
First, we fine-tune a LLaMA 3-8B model on a dataset of NL-STL pairs to obtain an initial STL generator.
% Second, we define four reward metrics that capture different aspects of STL quality: (1) template natural language similarity, (2) atomic proposition alignment, (3) formula conciseness, and (4) STL-level similarity. 
Second, for each metric, we train a separate reward model using preference-based paired examples and apply a curriculum learning strategy to improve the accuracy of the reward model feedback.
% For each metric, we train a separate reward model using preference-based paired examples. To enhance feedback accuracy, we apply a curriculum learning strategy, where training samples are introduced progressively from easier to harder based on the difficulty specific to each metric.
% , we design four metrics as reinforcement learning feedback: template natural language similarity, atomic proposition alignment, formula conciseness, and STL-level simnilarity. 
% Secondly, inspired by RLAIF, we train separate reward models for each feedback dimension using preference-based paired examples. To improve reward accuracy and training stability, a difficulty-aware curriculum learning strategy is applied during reward model training: for each reward, training samples are ranked by metric-specific difficulty and examples are introduced progressively from easier to harder.
Finally, the reward outputs from all models are aggregated into a unified scalar signal that provides feedback for PPO to optimize the STL generator.
% Secondly, inspired by RLAIF, we train separate reward models on corresponding paired datasets and combine the reward output by each reward model to provide feedback for PPO training. Additionally, we incorporate curriculum learning by sorting training examples based on the difficulty of the STL formulas, allowing the model to progressively learn from simpler to more complex cases.
% In what follows, we first introduce the details in each step, and then we summarize the PPO optimization process of RESTL.
% as Algorithm~\ref{alg:stl-ppo} in~\ref{subsec:whole_process}.
%\vspace{-0.2cm}

% \begin{figure}[htbp]
%     \centering
%     % 第一个子图
%     \subfloat[First Subfigure Caption]{%
%         \includegraphics[width=0.24\textwidth]{pictures/framework_1.png}%
%         \label{fig:sub1}
%     }\hspace{0.05cm}  % 调整子图之间的水平间隔
%     % 第二个子图
%     \subfloat[Second Subfigure Caption]{%
%         \includegraphics[width=0.24\textwidth]{pictures/framework_2.png}%
%         \label{fig:sub2}
%     }
%     \caption{Main Figure Caption with Two Subfigures}
%     \label{fig:main}
% \end{figure}

\subsection{STL Generator Initialization} \label{subsec:stl_initial_generator}

We fine-tune the LLaMA 3-8B model on a dataset of NL-STL pairs to obtain an initial STL generator capable of transforming natural language into STL. 
% This equips the model with foundational NL-to-STL transformation capabilities.
% Each NL-STL pair in the dataset is structured with three attributes: \texttt{Instruction}, \texttt{Input}, and \texttt{Output}. 
% Each NL-STL pair in the dataset is structured with three attributes: 
% the \texttt{Instruction} provides guidance for transforming, the \texttt{Input} NL description, and the target \texttt{Output} STL formula.
Each pair includes an \texttt{Instruction} guiding the transformation, an \texttt{Input} NL description, and the target \texttt{Output} STL formula. The model learns to generate the STL formula from the input and instruction.
% The \texttt{Instruction} provides detailed guidance on transforming the natural language description into a corresponding STL formula. The \texttt{Input} is the natural language description to be transformed and is input to the LLM along with the \texttt{Instruction}. The LLM is trained to generate the STL formula corresponding to the given \texttt{Input} as the \texttt{Output}. 

After training, the model serves as the initial STL generator with basic NL-to-STL transformation capabilities, ready for further fine-tuning on downstream tasks.
% which can be further adapted to downstream tasks.

\subsection{Multi-aspect Metric}\label{subsec:feedbacks}
\begin{figure*}
    \centering
    \includegraphics[width=1\linewidth]{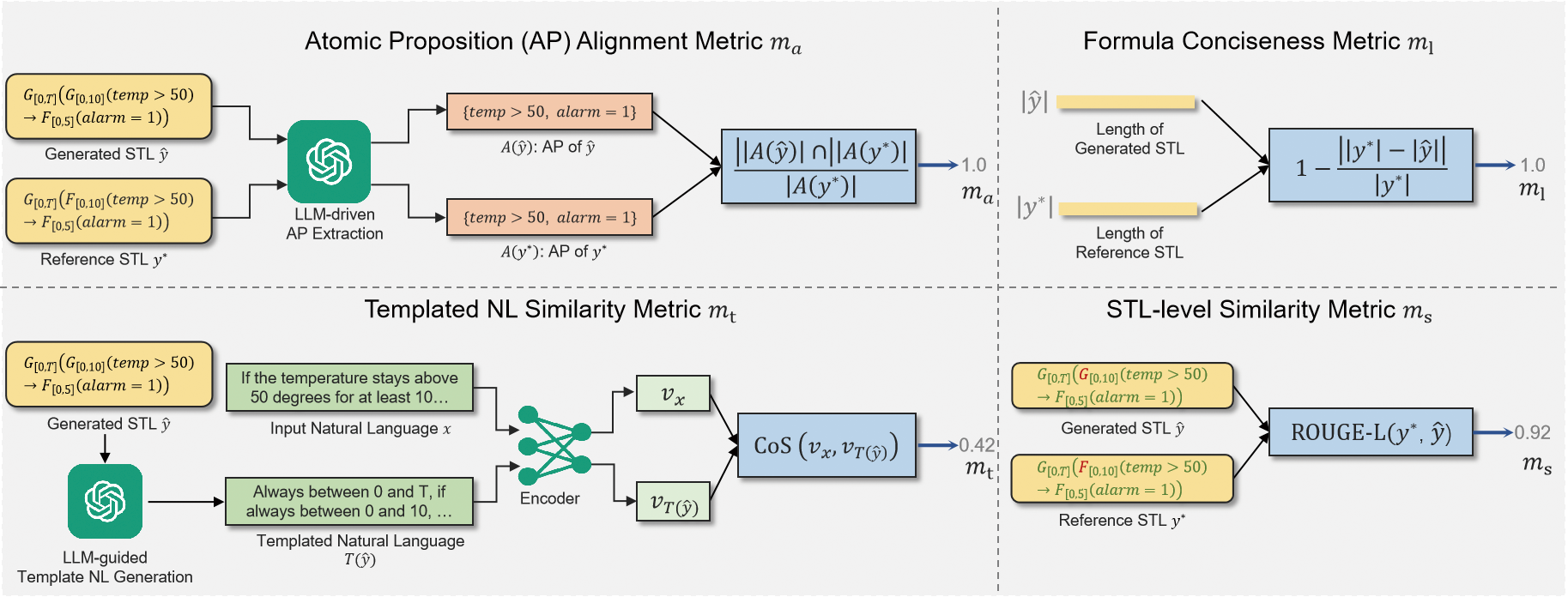}
    \caption{Design of the four reward metrics $m_{\text{a}}$, $m_{\text{t}}$, $m_{\text{l}}$, and $m_{\text{s}}$. We present the evaluation process of the same natural language input and its corresponding generated STL formula under different reward metrics.}
    \label{fig:rewardmetric}
\end{figure*}

% To address the common issues encountered when translating natural language into STL formulas using LLMs—such as misused temporal operators, incorrect atomic propositions, and overly complex expressions we design four reward functions as RL feedback.
% These reward functions help constrain the model’s generation process to accurate and interpretable STL generation.
To address common issues in transforming natural language into STL formulas using LLMs, 
% such as the misuse of temporal operators, incorrect atomic propositions, and overly complex expressions, 
we design four reward functions as reinforcement learning feedback as shown in Figure~\ref{fig:rewardmetric}. These reward functions guide the model toward generating more accurate STL formulas.

\subsubsection{Atomic Proposition Alignment} A common issue in NL-to-STL transformation is the incorrect identification or omission of atomic propositions, which are the fundamental variables in the formula. To evaluate the alignment between the generated formula \(\hat{y}\) and the ground truth \(y^*\), we extract their atomic propositions \(A(\hat{y})\) and \(A(y^*)\) respectively using LLMs, then compute the precision metric: \( m_{\text{a}} = |A(\hat{y}) \cap A(y^*)| / |A(y^*)| \).

% \[
% m_{\text{a}} = \frac{| A(\hat{y}) \cap A(y^*)|}{|A(y^*)|}.
% \]
% A common issue in STL generation is the incorrect interpretation of variables, thresholds, or time ranges in atomic propositions. To assess the alignment of atomic propositions between the generated formula \(\hat{y}\) and the ground truth \(y\), we extract the atomic propositions \(A(\hat{y})\) and \(A(y)\) from both and compute the precision-based matching score, defined as the metric \(m_{\text{AP}}\):
% \[
% m_{\text{AP}} = \frac{|A(\hat{y}) \cap A(y)|}{|A(y)|}
% \]
% This metric measures the degree of alignment of key proposition elements, ensuring local correctness in the formula.

\subsubsection{Templated NL Similarity} LLMs often produce semantic errors in NL-to-STL transformation, such as incorrect temporal logic operators, thresholds, or time ranges, which cause mismatches with the original intention of natural language. To enforce consistency at the semantic level, we reverse-map the generated STL formula \(\hat{y}\) into a templated natural language \(\text{T}(\hat{y})\). 
These templated natural language sentences are generated by LLMs based on templates defined by STL syntax rules.
We then use a pre-trained language model encoder (e.g., BERT) to convert both \(\text{T}(\hat{y})\) and the original input \(x\) into dense vector representations $v_{\text{T}(\hat{y})}$ and \(v_x\). Their semantic similarity is computed by cosine similarity as the metric \(m_{\text{t}} = \text{CoS}(v_x, v_{\text{T}(\hat{y})})\). 
% \[
% m_{\text{t}} = \text{CoS}(v_x, v_{\text{T}(\hat{y})}).
% \]
% LLMs often make mistakes in selecting appropriate temporal logic operators when generating STL formulas, leading to semantic mismatches with the original user intent. To enforce consistency at the operator level, we perform a reverse mapping of the generated STL formula \(\hat{y}\) into a structured natural language template \(T(\hat{y})\), and compute the semantic similarity between this template and the original natural language input \(x\), as the metric \(m_{\text{NL}}\). We use a pre-trained language model encoder (such as BERT) to extract dense representations \(v_T\) and \(v_x\) of \(T(\hat{y})\) and \(x\), respectively, and compute the cosine similarity:
% \[
% m_{\text{NL}} = \text{CS}(v_T, v_x)
% \]
% This metric encourages the model to maintain consistency with the original intent in operator selection.

\subsubsection{Formula Succinctness} To encourage the generation of concise and human-readable STL formulas, we design a reward function based on length difference. Let \(|\hat{y}|\) and \(|y^*|\) denote the number of characters in the generated STL formula and the ground truth formula, respectively. We define the normalized length difference metric \(m_{\text{l}}\) as:
\( m_{\text{l}} = 1 - |\,|y^*| - |\hat{y}|\,| / |y^*| \)
% \[
% m_{\text{l}} = 1 - \frac{|\  |y^*|-|\hat{y}|\ |}{|y^*|}.
% \]
This metric rewards formulas close in length to the reference. Formulas that are too long may include redundant or unnecessary components, while formulas that are too short may omit important logical content. Formulas with length near the reference get higher scores.

% To encourage the model to generate more concise and human-readable STL formulas, we design a reward function based on length difference. Let \(|\hat{y}|\) and \(|y|\) represent the lengths (in terms of tokens or symbols) of the generated and ground truth formulas, respectively. We define the normalized length difference metric \(m_{\text{len}}\) as:
% \[
% m_{\text{len}} = 1 - \frac{|\ |\hat{y}| - |y|\ |}{|y|}
% \]
% This metric rewards shorter formulas that are closer in length to the ground truth, reducing redundancy and the likelihood of errors.

\subsubsection{STL-level Similarity} To measure the overall similarity between the generated STL formula and the ground truth, we adopt the ROUGE-L~\cite{lin2004rouge} score as the metric \(m_{\text{r}}\). ROUGE-L evaluates the longest common subsequence between two sequences, capturing both content overlap and order information. This makes it suitable for assessing the structural and semantic consistency of STL formulas.
% , which are sensitive to the sequence and arrangement of operators and propositions.
Given the generated formula \(\hat{y}\) and ground truth \(y^*\), the metric is defined as:
\(
m_{\text{s}} = \text{ROUGE-L}(y^*,\hat{y}).
\)

\subsection{Reward Model Training} \label{subsec:reward_model_training}

% Without a reward model, the end-to-end RL approach for generating STL formulas requires a large number of training steps, we introduce reward models that evaluate the generated STL formulas, effectively decoupling the evaluation process from the STL generation and enabling more efficient training. 
% Building upon the ideas of RLAIF, we train four distinct reward models based on the respective metrics for the STL generation task. 
% These reward models are based on the LLaMA 3-8B model, which contains far fewer parameters than the generation model. 
% During RL training, rewards are computed using the outputs of these reward models instead of from the complete generation process, thereby reducing both per-iteration processing time and overall training duration.

We introduce reward models based on LLaMA 3-8B to evaluate generated STL formulas.
For a given input $x$, the initial STL generator produces multiple candidate formulas $\hat{y}_1, \dots, \hat{y}_k$ selected by minimizing pairwise ROUGE similarity, where $k$ is a hyperparameter.
% For each reward metric, we compute scores for the candidates and compare them to construct (`chosen', `rejected') pairs, where the higher-scoring candidate is labeled as `chosen'. This process generates training data that reflects relative quality judgments across candidates, enabling reward models to learn fine-grained preferences for STL generation quality.
All candidate formulas are evaluated with the four reward functions based on the metrics \( m_{\text{a}} \), \( m_{\text{t}} \), \( m_{\text{l}} \), and \( m_{\text{s}} \). For each reward metric, we compute scores for the candidates and compare them to construct (`chosen', `rejected') pairs, where the higher-scoring candidate is labeled as `chosen'. For example, for the metric \( m_{\text{s}} \), if \( m_{\text{s}}(\hat{y}^{(i)}) > m_{\text{s}}(\hat{y}^{(j)}) \), we collect the following pair:
{\footnotesize
\[
\left\{ (\text{chosen}: [x, \hat{y}^{(i)}],\ \text{rejected}: [x, \hat{y}^{(j)}])\ \middle|\ m_{\text{s}}(\hat{y}^{(i)}) > m_{\text{s}}(\hat{y}^{(j)}) \right\}.
\]
}

% This process generates preference datasets that reflect relative quality judgments across candidates, enabling reward models to learn fine-grained preferences for STL generation quality.
This process creates preference data that enables reward models to learn fine-grained preferences for STL generation quality.

When training the reward models, we use the Bradley-Terry model~\cite{bradley1952rank} to formulate the preference distribution with the reward model \( r_{\psi} \) as follows:
\[
P_{\psi}(y_c \succ y_r | x) = \sigma(r_{\psi}(x, y_c) - r_{\psi}(x, y_r)),
\]
where \( \sigma \) is the logistic function, \( y_c \) denotes the chosen STL, and \( y_r \) represents the rejected STL. This is treated as a binary classification task, yielding the following negative log-likelihood loss function:
\[
L_{\text{rm}} = -\mathbb{E}_{D_{\text{rm}}} \left[ \log \sigma(r_{\psi}(x, y_c) - r_{\psi}(x, y_r)) \right],
\]
where \( D_{\text{rm}} \) is the preference dataset.

In this work, we initialize the reward models using the initial STL generator. Additionally, a linear layer is added on top of the final transformer layer to produce a scalar prediction representing the reward value. 

Let \( r_{\text{a}} \), \( r_{\text{t}} \), \( r_{\text{l}} \), and \( r_{\text{s}} \) denote the \textbf{reward models} for the metrics \( m_{\text{a}} \), \( m_{\text{t}} \), \( m_{\text{l}} \), and \( m_{\text{s}} \), respectively. Given the input natural language \( x \) and the generated STL formula \( \hat{y} \), the corresponding rewards can be computed as \( r_{\text{a}}(x, \hat{y}) \), \( r_{\text{t} }(x, \hat{y}) \), \( r_{\text{l}}(x, \hat{y}) \), and \( r_{\text{s}}(x, \hat{y}) \), abbreviated as \( r_{\text{a}}(\hat{y}) \), \( r_{\text{t}}(\hat{y}) \), \( r_{\text{l}}(\hat{y}) \), and \( r_{\text{s}}(\hat{y}) \).
To facilitate aggregation of the rewards, the scores from each reward model are scaled to the range [0, 1].

\subsection{Curriculum Learning for Reward Models} \label{subsec:curriculum_learning}

To improve the effectiveness of reward models, we incorporate curriculum learning by organizing training examples from easy to more challenging cases. We define separate curricula based on the four reward metrics, each capturing distinct reward model evaluation criteria. 
% This approach enables the reward model to achieve more accurate evaluation results.

\subsubsection{Atomic Proposition Curriculum for $\bm{r_{\text{a}}}$} This curriculum is used for training the atomic proposition alignment reward model. We count the number of atomic propositions in the ground truth STL formula and sort the training samples in ascending order of this count. Samples with fewer atomic propositions are considered easier and are prioritized, enabling the model can gradually learn to assess atomic proposition alignment.
% \[
% \text{Difficulty}_{\text{ap}} = \text{AtomsCount}(y^*).
% \]

\subsubsection{NL Similarity Curriculum for $ \bm{r_{\text{t}}} $} When training the reward model of templated natural language similarity, for each natural language description in the preference dataset, we first convert the corresponding three generated STL formulas into three templated natural language sentences. 
Then, we compute the cosine similarity between the original natural language sentence and each templated sentence, and use the average similarity of the three pairs to determine the sample order. 
% Then, we compute the semantic similarity between the original natural language and each templated natural language sentence, taking the average similarity of the three pairs as the basis for sample ordering. 
These training data are sorted in ascending order of difficulty scores.
% Training data are sorted in descending order of average templated NL similarity scores.
% , introducing samples with higher similarity earlier, followed by those with lower similarity.
% \[
% \text{Difficulty}_{\text{nl}} = 1 - \frac{1}{3} \sum_{i=1}^{3} \text{CoS}(x, \text{T}(\hat{y}_i)).
% \]

\subsubsection{STL Formula Length Curriculum for $\bm{r_{\text{l}}}$} For training the formula succinctness reward model, samples are sorted by the length of the ground truth STL formula. Shorter STL formulas are regarded as easier and introduced earlier, with progressively longer formulas introduced as training proceeds.
% \[
% \text{Difficulty}_{\text{len}} = \text{len}(y^*).
% \]

\subsubsection{STL Similarity Curriculum for $\bm{r_{\text{s}}}$} This curriculum is designed to train the STL-level similarity reward model. 
For each group of three generated STL formulas in the preference dataset, we first calculate the ROUGE-L score between each generated formula and the ground truth STL formula. 
The average of these three ROUGE-L scores is then used to compute the difficulty score.
Training samples are sorted in ascending order of difficulty scores.

\subsection{Reinforcement Learning for STL Generator} \label{subsec:whole_process}

% The RL training process of RESTL is shown in Algorithm~\ref{alg:stl-ppo}. It starts with the prepared initial STL generator \( G_{\theta_0} \) (see Section~\ref{subsec:stl_initial_generator}). 

%and iteratively updates the policy using PPO. For each training example, the generator samples multiple candidate STL formulas. Each candidate is evaluated with four reward models. These rewards are aggregated and used to compute the PPO loss with a KL penalty. Finally, the generator is updated using gradient descent.

% In each training epoch, we compute the aggregated multi-aspect reward for each sample $x^{(i)} \in D$ . First, for each NL description $x^{(i)}$, we generate a corresponding STL formula $\hat{y}^{(i)}$ . Then, we compute its scores using the four reward models and aggregate them into an overall reward. Formally, g
Given a natural language instruction \( x \) and the generated STL formula \( \hat{y} \), the overall reward is computed as:
\begin{equation*}
r_{\text{RL}}(\hat{y}) = \lambda_1 r_{\text{a}}(\hat{y}) + \lambda_2 r_{\text{t}}(\hat{y}) + \lambda_3 r_{\text{l}}(\hat{y}) + \lambda_4 r_{\text{s}}(\hat{y}),
% \label{eq:stl-reward}
\end{equation*}
% where \( r_{\text{a}} \), \( r_{\text{t}} \), \( r_{\text{l}} \), and \( r_{\text{s}} \) are the reward scores corresponding to the atomic proposition alignment $m_{\text{a}}$, templated NL similarity $m_{\text{t}}$, formula conciseness $m_{\text{l}}$, and STL-level similarity $m_{\text{s}}$, proposed in Section~\ref{subsec:feedbacks}, respectively. 
where the hyperparameters \( \lambda_1, \lambda_2, \lambda_3, \lambda_4 \) control the relative importance of each reward. 
The reward objective $r_{\text{total}}$ is to maximize expected reward while constraining deviation from the initial policy \( G_{\theta_0} \). 
%The objective is given by:
\begin{equation*}
r_{\text{total}} = r_{\text{RL}}(\hat{y}) - \eta \cdot \text{KL}(G_{\theta} \parallel G_{\theta_0}),
% \label{eq:ppo-loss}
\end{equation*}
where \( \eta \) is a KL penalty coefficient. This term stabilizes training by penalizing large shifts from the pre-trained generator.

% Finally, we apply the Proximal Policy Optimization (PPO) algorithm~\cite{schulman2017proximal} to optimize the STL generator \( G_{\theta} \) (Line~\ref{line:apply_ppo_start} to \ref{line:apply_ppo_end}). 
Finally, we apply the PPO algorithm~\cite{schulman2017proximal} to optimize the STL generator \( G_{\theta} \) using the KL-regularized reward signal. 
\section{Experiments} \label{sec:experiment}
In this section, we conduct comprehensive experiments to evaluate our proposed method RESTL.
% \footnote{RESTL project is available at: \url{https://zenodo.org/records/15554395}}, and answer the following four research questions.
% \begin{enumerate}
%     % \item {\bf RQ1:} How does RESTL perform compared to existing baselines on STL transformation across different datasets?
%     \item {\bf RQ1:} How does RESTL perform on STL transformation across different datasets? 
%     % \item {\bf RQ2:} What is the individual contribution of each reward signal in the multi-aspect reward model?
%     % \item {\bf RQ2:} How about the impacts of the multi-aspect metrics? 
%     \item {\bf RQ2:} How do the multi-aspect metrics impact the STL transformation? 
%     \item {\bf RQ3:} How does the curriculum learning impact the reward models and RESTL framework?
%     % \item {\bf RQ4:} How does reward model-based feedback compare to direct metric supervision in reinforcement learning for STL transformation?
%     % \item {\bf RQ4:} How does the feedback from reward models compare to direct metric supervision in reinforcement learning for STL transformation?
%     % \item {\bf RQ4:} Does the reward model provide better feedback than direct metric supervision in reinforcement learning for STL transformation?
%     \item {\bf RQ4:} Does the reward model offer superior feedback compared to direct metric supervision in reinforcement learning for STL transformation?
% \end{enumerate}

\subsection{Experimental Settings}
We first introduce the empirical settings, including datasets, baselines, evaluation measures, and implementation details.

\subsubsection{Datasets} 
We conduct experiments on two datasets for NL-to-STL transformation, including the DeepSTL dataset~\cite{HeBNIG22} and STL-DivEn dataset~\cite{fang2025enhancingtransformationnaturallanguage}.
The DeepSTL dataset is synthetically generated using a grammar-based generator that samples STL formulas from predefined templates and operator distributions. 
% The corresponding English descriptions are composed by mapping the structure of each formula to fixed linguistic templates in a rule-based manner.
STL-DivEn is created through a hybrid approach that integrates GPT-4-based generation and human verification.
% The diversity of NL-STL pairs is enhanced by guiding GPT-4 generation with clustering-based sample selection and applying diversity filtering.
% Semantic correctness is ensured by manual review of all generated samples.
For a fair comparison, we randomly select 14,000 samples from each dataset for training and 2,000 samples for testing.

\subsubsection{Baselines}
We compare RESTL with five baseline methods: DeepSTL~\cite{HeBNIG22}, GPT-3.5\footnote{https://platform.openai.com/docs/models/gpt-3-5-turbo}
, GPT-4\footnote{https://platform.openai.com/docs/models/gpt-4-turbo-and-gpt-4}, DeepSeek~\cite{guo2025deepseek}, and KGST~\cite{fang2025enhancingtransformationnaturallanguage}. 
In our implementation, we adopt the ``gpt-4-0125-preview'' version of GPT-4, the ``gpt-3.5-turbo-1106'' version of GPT-3.5, and the ``DeepSeek-V1'' version of DeepSeek.
DeepSTL adopts the Transformer architecture for training and optimizes model parameters using the Adam algorithm~\cite{kingma2014adam}.
KGST is the SOTA model, which adopts a two-stage architecture: it first fine-tunes a LLaMA 3-8B model on an NL-STL dataset to produce preliminary STL formulas. In the second stage, it retrieves the top-5 most similar NL-STL pairs from the training set as reference examples, and leverages GPT-4 to refine the initial outputs, generating the final STL formulas.

\subsubsection{Evaluation Measures}
% To assess the quality of STL generation, we employ both automatic evaluation metrics and human evaluation. 
% For automatic evaluation, following previous studies~\cite{HeBNIG22}, we adopt the accuracy of the STL formula, the accuracy of the template and BLEU~\cite{papineni2002bleu} as evaluation metrics.
% STL Formula Accuracy quantifies exact token-level matches between the generated and reference formulas, capturing the syntactic correctness. 
% Template Accuracy measures the alignment of the structural templates of two formulas, providing insight into whether the core logical form is preserved. 
% BLEU was originally designed for machine translation and evaluates n-gram overlap, reflecting lexical semantic similarity.
% The meaning of Formula and Template Accuracy are shown in appendix.
To evaluate STL generation quality, we use both automatic metrics and human evaluation. Following prior work~\cite{HeBNIG22}, we adopt formula accuracy, template accuracy, and BLEU~\cite{papineni2002bleu}. Formula accuracy measures exact token-level matches, template accuracy assesses structural alignment, and BLEU captures n-gram overlap for lexical similarity. Definitions of formula and template accuracy are provided in the Appendix C.

% \paragraph{STL Formula Accuracy} Let $y^* = [y_1^*, y_2^*, \ldots, y_n^*]$ be the reference token sequence and $\hat{y} = [\hat{y}_1, \hat{y}_2, \ldots, \hat{y}_n]$ be the predicted sequence of the same length. The formula accuracy is computed as:
% \[
% A_F = \frac{1}{n} \sum_{i=1}^{n} \mathbb{I}[y_i^* = \hat{y}_i].
% \]
% \paragraph{Template Accuracy} Let $\text{Tem}(y^*)$ and $\text{Tem}(\hat{y})$ denote the abstracted STL templates of the reference and predicted formulas, respectively, each consisting of $m$ tokens. The template accuracy is defined similarly:
% \[
% A_\text{T} = \frac{1}{m} \sum_{j=1}^{m} \mathbb{I}[\text{Tem}(y_j^*) = \text{Tem}(\hat{y}_j)].
% \]

% The following example shows both the formula and its corresponding template for the reference and predicted outputs.
% \[
% \scalebox{0.85}{$
% \begin{aligned}
% \text{Formula:} \; & \G ( x > 10 ) \rightarrow \F ( y < 3 )
% \Rightarrow \, \text{Template:} \;  \G ( \phi ) \rightarrow \F ( \phi ) \\
% \text{Formula:} \; & \G ( x > 10 ) \rightarrow \F ( z < 3 )  
% \Rightarrow \, \text{Template:} \;  \G ( \phi ) \rightarrow \F ( \phi )
% \end{aligned}
% $}
% \]
% The formula contains 9 tokens. All tokens coincide except for the variable ``y'' vs ``z'' in the atomic proposition, resulting in $A_F = \frac{8}{9}$. When both formulas are converted to templates by replacing atomic propositions with placeholders (e.g., $\phi$), all tokens align perfectly: $A_\text{T} = 1$.

For human evaluation, we randomly sample 100 NL-STL pairs from testing sets of STL-DivEn and DeepSTL. Five trained annotators, familiar with STL semantics and syntax, assess the output without knowing the model source. 
The evaluation is based on three criteria: readability (i.e. ease of understanding), syntactic correctness, and consistency with
the original semantics. 
Readability is judged only if the first two criteria are met. 
This evaluation is designed to complement metric-based methods, which may miss cases where STL formulas differ in form but share the same meaning. 
Each comparison between RESTL and a baseline is labeled as win, loss, or tie, reflecting overall clarity and correctness. 
% For human evaluation, we randomly sample 100 NL-STL pairs from testing sets of STL-DivEn and DeepSTL. Five trained annotators, familiar with STL semantics and syntax, assess the output without knowing the model source. The evaluation is based on three criteria: readability (i.e. ease of understanding), syntactic correctness, and consistency with the original semantics. 
% Each comparison between RESTL and a baseline is labeled as win, loss, or tie, reflecting overall clarity and correctness.
% Each comparison between RESTL and a baseline output is labeled as win, loss, or tie, indicating whether RESTL's output is judged to be better, worse, or equivalent in terms of overall clarity and correctness.

\subsubsection{Implementation Details}
Our experiments run on 8 NVIDIA GeForce RTX 4090 GPUs (24GB each). Each reward model is fine-tuned on LLaMA 3-8B with a linear value head for 5 epochs using Adam (lr=5e-5, batch=16). The STL generator is fine-tuned via PPO for 80,000 steps (batch=32, lr=1.41e-5, KL penalty $\eta=0.05$). The combined reward uses weighted scores with $\lambda_1=0.2$, $\lambda_2=0.25$, $\lambda_3=0.35$, and $\lambda_4=0.2$. More details, including hyperparameter discussions, are in Appendix D.

\subsection{Main Results}
% In the following, we evaluate RESTL through comprehensive experiments on two datasets: STL-DivEn and DeepSTL benchmarks and compare RESTL with the baselines through both automated metrics and human evaluation.
We demonstrate results on two datasets to evaluate RESTL. 

% \subsubsection{\textbf{RQ1:} How does RESTL perform compared to existing baselines on STL generation tasks across different datasets?}

% perform compared to existing baselines on STL generation tasks across different datasets?}

\begin{table}[!t]
    \centering    
    % Subtable for STL-DivEn
    \begin{subtable}[t]{\linewidth}
        \centering
        \caption{STL-DivEn dataset}
        \label{tab:stl_diven_results}
        % \rowcolors{2}{gray!10}{white}
        \resizebox{\linewidth}{!}{
        \begin{tabular}{lccc}
            \toprule
            \textbf{Model} & \textbf{Formula Acc.} & \textbf{Template Acc.} & \textbf{BLEU} \\
            \midrule
            DeepSTL & 0.1986 & 0.1883 & 0.0293 \\
            GPT-3.5 & 0.3018 & 0.3034 & 0.0424 \\
            GPT-4 & 0.4733 & 0.4741 & 0.0831 \\
            DeepSeek & 0.4790 & 0.4825 & 0.0791 \\
            KGST & 0.5587 & 0.5627 & 0.2142 \\
            RESTL (Ours) & \textbf{0.6838} & \textbf{0.6974} & \textbf{0.3347} \\
            \bottomrule
        \end{tabular}
        }
    \end{subtable}

    % \vspace{1em}  % Add vertical space between subtables

    % Subtable for DeepSTL
    \begin{subtable}[t]{\linewidth}
        \centering
        \caption{DeepSTL dataset}
        \label{tab:deepstl_results}
        % \rowcolors{2}{gray!10}{white}
        \resizebox{\linewidth}{!}{
        \begin{tabular}{lccc}
            \toprule
            \textbf{Model} & \textbf{Formula Acc.} & \textbf{Template Acc.} & \textbf{BLEU} \\
            \midrule
            DeepSTL & 0.2002 & 0.2916 & 0.3332 \\
            GPT-3.5 & 0.2145 & 0.3002 & 0.2249 \\
            GPT-4 & 0.2262 & 0.3048 & 0.2881 \\
            DeepSeek & 0.2537 & 0.3254 & 0.3982 \\
            KGST & 0.4538 & 0.4939 & 0.5686 \\
            RESTL (Ours) & \textbf{0.5985} & \textbf{0.6327} & \textbf{0.6783} \\
            \bottomrule
        \end{tabular}
        }
    \end{subtable}
    \caption{Performance comparison of RESTL and baselines on STL-DivEn and DeepSTL datasets.}
    \label{tab:main_results}
\end{table}

% \begin{table}[!t]
%     \centering
%     \caption{Performance comparison on STL-DivEn dataset.}
%     \label{tab:stl_diven_results}
%     \rowcolors{2}{gray!10}{white}
%     \begin{tabular}{lccc}
%         \toprule
%         \textbf{Model} & \textbf{Formula Acc.} & \textbf{Template Acc.} & \textbf{BLEU} \\
%         \midrule
%         DeepSTL & 0.1986 & 0.1883 & 0.0293 \\
%         GPT-3.5 & 0.3018 & 0.3034 & 0.0424 \\
%         GPT-4 & 0.4733 & 0.4741 & 0.0831 \\
%         DeepSeek & 0.4790 & 0.4825 & 0.0791 \\
%         KGST & 0.5587 & 0.5627 & 0.2142 \\
%         RESTL (Ours) & \textbf{0.5836} & \textbf{0.5874} & \textbf{0.2542} \\
%         \bottomrule
%     \end{tabular}
% \end{table}

% \begin{table}[!t]
%     \centering
%     \caption{Performance comparison on DeepSTL dataset.}
%     \label{tab:deepstl_results}
%     \rowcolors{2}{gray!10}{white}
%     \begin{tabular}{lccc}
%         \toprule
%         \textbf{Model} & \textbf{Formula Acc.} & \textbf{Template Acc.} & \textbf{BLEU} \\
%         \midrule
%         DeepSTL & 0.2002 & 0.2916 & 0.3332 \\
%         GPT-3.5 & 0.2145 & 0.3002 & 0.2249 \\
%         GPT-4 & 0.2262 & 0.3048 & 0.2881 \\
%         DeepSeek & 0.2537 & 0.3254 & 0.3982 \\
%         KGST & 0.4538 & 0.4939 & 0.5686 \\
%         RESTL (Ours) & \textbf{0.4952} & \textbf{0.5264} & \textbf{0.5931} \\
%         \bottomrule
%     \end{tabular}
% \end{table}

\begin{table}[!t]
\centering
\scalebox{0.85}{
\begin{tabular}{lcccccc}
\toprule
\multirow{2}{*}{Model} 
& \multicolumn{3}{c}{vs. STL-DivEn} 
& \multicolumn{3}{c}{vs. DeepSTL} \\
\cmidrule(lr){2-4} \cmidrule(lr){5-7}
& Win & Loss & Tie & Win & Loss & Tie \\
\midrule
DeepSeek & 64.2 & 12.3 & 23.5 & 56.3 & 17.2 & 26.5 \\
GPT-4    & 61.0 & 13.5 & 25.5 & 54.7 & 18.6 & 26.7 \\
KGST     & 58.7 & 15.9 & 25.4 & 52.8 & 19.7 & 27.5 \\
\bottomrule
\end{tabular}
}
\caption{Human evaluation (\%) of RESTL vs. baselines.}
\label{tb:humaneval-merged}
\end{table}

\subsubsection{Metric-Based Evaluation}
As shown in Table~\ref{tab:stl_diven_results}, on the STL-DivEn dataset, RESTL reaches a formula accuracy of 68.38\%, exceeding the strongest baseline KGST (55.87\%). 
% by approximately 5 percentage points. 
It also achieves a template accuracy of 69.74\% (vs. KGST’s 56.27\%) and a BLEU score of 0.3347, higher than KGST’s 0.2142. These results demonstrate RESTL’s superior ability in generating more accurate STL formulas.
 % syntactically and semantically accurate.
Similarly, as shown in Table~\ref{tab:deepstl_results}, RESTL achieves the best performance on the DeepSTL dataset, with a formula accuracy of 59.85\%, a template accuracy of 63.27\%, and a BLEU score of 0.6783. Compared to KGST (45.38\%, 49.39\%, and 0.5686, respectively), RESTL demonstrates clear improvements across all metrics.
We conduct a significance test, confirming that RESTL significantly outperforms existing models across datasets and metrics, i.e., $\text{p-value}<0.01$. Experimental results including measures of variability are provided in Appendix E.1.
% and generalization capability.

\subsubsection{Human Evaluation}
The results shown in Table~\ref{tb:humaneval-merged} indicate that annotators consistently prefer formulas generated by RESTL, as they exhibit more concise and readable expressions while maintaining semantic consistency. 
For example, RESTL achieved win rates of 64.2\%, 61.0\%, and 58.7\% against DeepSeek, GPT-4, and KGST, respectively. 
We attribute this readability advantage to the introduction of the formula succinctness reward ($m_{\text{l}}$) during reinforcement learning, which explicitly encourages the model to generate STL formulas with lengths close to the reference, thereby improving clarity and readability.
\subsection{Ablation Study}
% \subsubsection{\textbf{RQ2:} What is the individual contribution of each reward signal in the multi-aspect reward model?}

\begin{table}[!t]
\centering
% \caption{Performance comparison using different reward signals for RL training on STL-DivEn and DeepSTL datasets.}
\begin{subtable}[t]{1\linewidth}
    \centering
    \caption{STL-DivEn dataset}
    \resizebox{\linewidth}{!}{
    \begin{tabular}{lcccc}
    \toprule
    \textbf{Model} & \textbf{Formula Acc.} & \textbf{Template Acc.} & \textbf{BLEU}\\
    \midrule
    RESTL      & \textbf{0.6838} & \textbf{0.6974} & \textbf{0.3347}\\  
    - w/o $m_{\text{a}}$   & 0.6573 & 0.6594 & 0.3117 \\
    - w/o $m_{\text{t}}$       & 0.6424 & 0.6503 & 0.3095 \\
    - w/o $m_{\text{l}}$       & 0.6642 & 0.6709 & 0.3294  \\
    - w/o $m_{\text{s}}$       & 0.6314 & 0.6393 & 0.2913   \\
    \midrule
    LLaMA3 (Fine-tuned) & 0.4956 & 0.5007 & 0.1784 \\
    \bottomrule
    \end{tabular}
    }
\end{subtable}
% \vspace{2mm}

\begin{subtable}[t]{1\linewidth}
    \centering
    \caption{DeepSTL dataset}
    \resizebox{\linewidth}{!}{
    \begin{tabular}{lccc}
    \toprule
    \textbf{Model} & \textbf{Formula Acc.} & \textbf{Template Acc.} & \textbf{BLEU} \\
    \midrule
    RESTL      & \textbf{0.5985} & \textbf{0.6327} & \textbf{0.6783}\\
    - w/o $m_{\text{a}}$   & 0.5571 & 0.5642 & 0.6129 \\
    - w/o $m_{\text{t}}$       & 0.5683 & 0.5782 & 0.6293 \\
    - w/o $m_{\text{l}}$       & 0.5832 & 0.5927 & 0.6473  \\
    - w/o $m_{\text{s}}$       & 0.5409 & 0.5503 & 0.6091 \\
    \midrule
    LLaMA3 (Fine-tuned) & 0.2850 & 0.3285 & 0.5579 \\
    \bottomrule
    \end{tabular}
    }
\end{subtable}
\caption{Ablation study of different reward feedback on STL-DivEn and DeepSTL datasets.}
\label{tab:reward_rl_comparison}
\end{table}

% We conduct reinforcement learning training on two datasets using a single reward model or metric as feedback, and compare the results against those of the initial STL generator. 
% we conduct ablation studies on both datasets. In each variant, one reward metric is removed and the model is trained using the remaining three. We compare the results with those of the full RESTL model and the fine-tuned LLaMA3 baseline.
We conduct ablation studies on both datasets by removing one reward metric at a time and training the model with the remaining three. 
% We compare the results with those of the full RESTL model and the fine-tuned LLaMA 3-8B baseline. 
The results are shown in Table~\ref{tab:reward_rl_comparison}, and more details are provided in the Appendix E.2.
% Table~\ref{tab:reward_rl_comparison} presents the impact of four different metrics on the effectiveness of reinforcement learning for STL generation.

% To quantify the impact of each reward component in ReSTL, we conduct a detailed \textit{ablation study}, where we systematically disable one reward signal at a time and observe the change in STL generation performance.

(1) Impact of $m_{\text{a}}$ for atomic proposition alignment: for STL-DivEn dataset, removing $m_{\text{a}}$ yields 65.73\% formula accuracy, 65.94\% template accuracy, and a BLEU score of 0.3117, all outperforming the fine-tuned LLaMA 3-8B baseline. For DeepSTL dataset, removing $m_{\text{a}}$ still better than baseline. However, these improvements remain limited compared to the full multi-reward combination in RESTL. These results indicate that $m_{\text{a}}$ as a reward feedback is effective. 

(2) Impact of $m_{\text{t}}$ for templated NL similarity: 
% Without $m_{\text{t}}$, formula accuracy drops to 64.24\% on STL-DivEn and 56.83\% on DeepSTL, template accuracy of 65.03\% and 57.83\% with BLEU scores of 0.3095 and 0.6293, respectively, showing $m_{\text{t}}$ is a useful reward.
without $m_{\text{t}}$, formula accuracy drops to 64.24\% on STL-DivEn and 56.83\% on DeepSTL, with template accuracies of 65.03\% and 57.82\% and BLEU scores of 0.3095 and 0.6293, respectively, showing that $m_{\text{t}}$ is effective.

% On STL-DivEn, training without $m_{\text{t}}$ results in a formula accuracy of 58.72\% and a BLEU score of 0.2365. On DeepSTL, it achieves a formula accuracy of 49.92\% and a BLEU score of 0.5913. Both results are higher than those of the fine-tuned LLaMA 3-8B baseline but lower than the full RESTL model, demonstrating the effectiveness of $m_{\text{t}}$ as a reward signal.

(3) Impact of $m_{\text{l}}$ for formula succinctness: 
removing $m_{\text{l}}$ yields the highest automatic scores among all ablation settings, with 66.42\% formula accuracy and a BLEU score of 0.3294 on STL-DivEn, and 58.32\% formula accuracy and 0.6473 BLEU on DeepSTL, 
% indicating that while $m_{\text{l}}$ has limited impact on metrics, it still helps improve formula readability.
indicating that $m_{\text{l}}$ has limited metric impact but improves readability.

% On STL-DivEn, without formula succinct reward $m_{\text{l}}$, it leads to a formula accuracy of 59.13\% and a BLEU score of 0.2417, which are the highest among all the ablation settings.
% On DeepSTL, the formula accuracy achieves 50.21\%, the BLEU score remains at 0.5977. 
% These experimental results show that while $m_{\text{l}}$ encourages generating more concise STL formulas, its contribution to improving automatic metrics is relatively limited.
% Nonetheless, its auxiliary role remains valuable for enhancing readability and clarity.

(4) Impact of $m_{\text{s}}$ for STL-level similarity: 
removing $m_{\text{s}}$ leads to the largest performance drop among all ablations, with 63.14\% formula accuracy and a BLEU score of 0.2913 on STL-DivEn, and 54.09\% accuracy and 0.6091 BLEU on DeepSTL, 
% showing the largest performance drop among all ablations and highlighting $m_{\text{s}}$ as the most important reward for capturing STL-level fidelity.
showing that $m_{\text{s}}$ is the most significant reward for improving accuracy.

% On STL-DivEn, removing $m_{\text{s}}$ results in the lowest performance among four ablation settings, with a formula accuracy of 58.03\% and a BLEU score of 0.2319. 
% On the DeepSTL, it achieves 49.02\% formula accuracy and BLEU score of 0.5827, outperforming the fine-tuned LLaMA 3-8B baseline and still falling short of the full RESTL model. 
% These experimental results confirm that $m_{\text{s}}$ is the most effective feedback in the ReSTL framework, as it directly captures the similarity between the generated STL formula and the reference.
% effectively captures global structural similarity and contributes substantially to improving formula quality, especially when used in conjunction with other reward signals.

\subsection{Error Analysis}

\begin{table}[!t]
\centering
\resizebox{1\linewidth}{!}{
\begin{tabular}{lcccc}
\toprule
\multicolumn{1}{c}{Model} & \begin{tabular}[c]{@{}c@{}}\#AP \end{tabular} & \begin{tabular}[c]{@{}c@{}}\#Operator\end{tabular} & \begin{tabular}[c]{@{}c@{}}\#Value\end{tabular} & \begin{tabular}[c]{@{}c@{}}\#Redundancy \end{tabular}\\ 
\midrule
LLaMA3-8B (Fine-tuned)            &  19     &    39  &    27   &  22  \\  
KGST          &     15 &    27  &   25  &   26  \\
RESTL        &     6  &    15  &   18  &   14   \\ 
\bottomrule
\end{tabular}
}
%\vspace{-5mm}
\caption{\label{tb:erroranalysis}Error analysis of RESTL, KGST, and fine-tuned LLaMA3-8B. \#AP, \#Operator, \#Value, and \#Redundancy denote counts of atomic proposition, operator, value, and redundancy errors, respectively.}
\end{table}
% \caption{\label{tb:erroranalysis}Error type analysis of RESTL, KGST and LLaMA3-8B (Fine-tuned). \#AP, \#Operator, \#Value, and \#Redundancy represent the counts of atomic proposition errors, operator misuse, value errors, and redundancy errors, respectively.}
% \end{table}

As shown in Table~\ref{tb:erroranalysis}, we analyze 100 STL formulas generated by RESTL, KGST, and fine-tuned LLaMA 3-8B, categorizing errors into four types: atomic proposition (AP), operator, numerical value, and redundancy. Compared to the baselines, RESTL shows fewer AP errors due to the AP Alignment reward, and fewer operator and value errors thanks to the Templated NL Similarity metric. The Formula Conciseness reward also helps reduce redundancy, whereas KGST tends to include more irrelevant content, likely due to its retrieval-augmented design. The STL-level Similarity metric is excluded from this analysis as it serves as a global training signal.

\subsection{Impacts of Curriculum Learning}

% \begin{table}[t]
% \centering
% \caption{Ablation study of curriculum learning strategies on STL-DivEn.}
% \label{tab:curriculum_ablation}
% \scalebox{0.95}{
% \begin{tabular}{lccc}
% \toprule
% \textbf{Curriculum Setting} & \textbf{Formula Acc.} & \textbf{Template Acc.} & \textbf{BLEU} \\
% \midrule
% Full CL              & \textbf{0.5836} & \textbf{0.5874} & \textbf{0.2542} \\
% No-StructCL             & 0.5586 & 0.5712 & 0.2319 \\
% No-AlignCL              & 0.5623 & 0.5761 & 0.2327 \\
% No-LengthCL             & 0.5665 & 0.5726 & 0.2334 \\
% No-ROUGECL              & 0.5534 & 0.5614 & 0.2305 \\
% No Curriculum           & 0.5346 & 0.5307 & 0.2193 \\
% \bottomrule
% \end{tabular}
% }
% \end{table}

\begin{figure}[!t]
    \centering
    \includegraphics[scale=0.4]{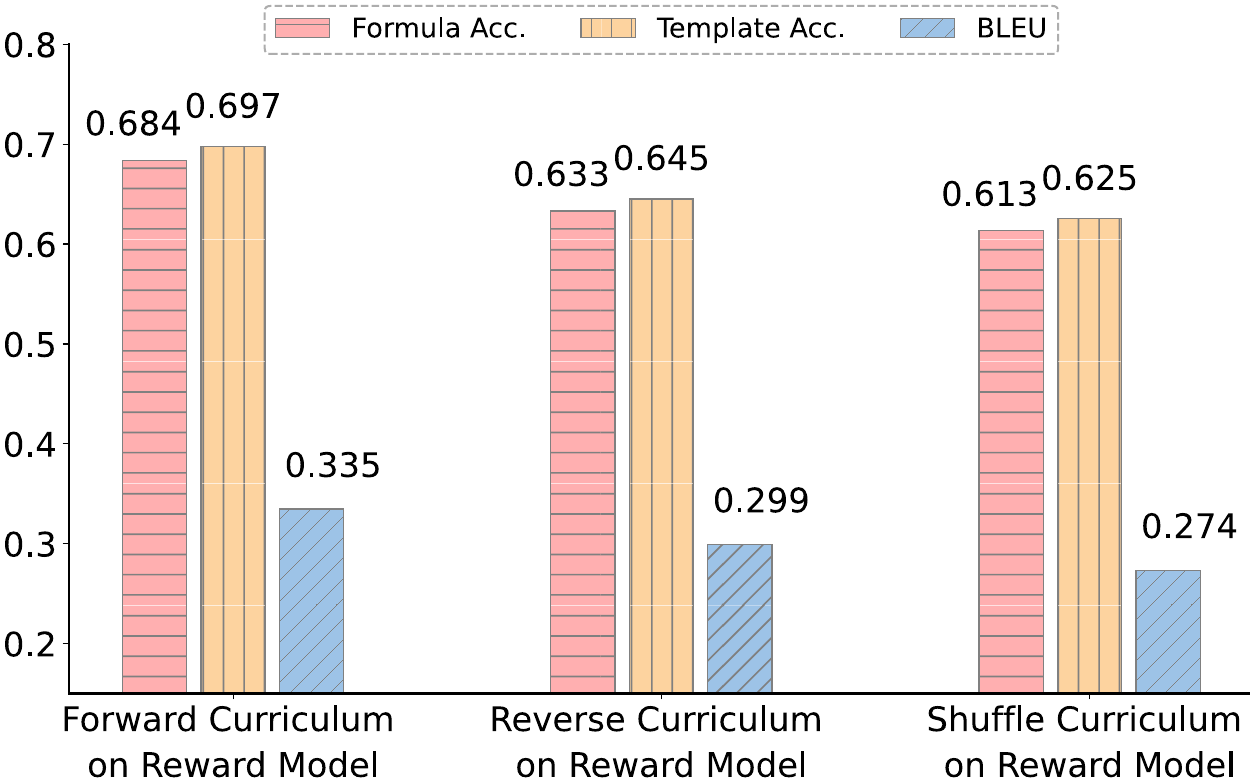}
    \caption{Scheduling strategy impact of different curricula.}
    \label{fig:curriculum_learning}
\end{figure}

% As shown in Figure~\ref{fig:curriculum_learning}, we compare how different training methods for reward models affect the performance of the RESTL framework. 
As shown in Figure~\ref{fig:curriculum_learning}, we compare how different training data orders for reward models affect the performance of the RESTL framework.
The results show that reward models trained with curriculum learning more effectively improve formula accuracy, template accuracy, and BLEU scores, significantly enhancing the overall performance of RESTL. Detailed impacts on individual reward models are provided in Appendix E.3.

\subsection{Reward Model vs. Direct Metric in RL}

\begin{figure}[!t]
    \centering
    \includegraphics[width=\linewidth]{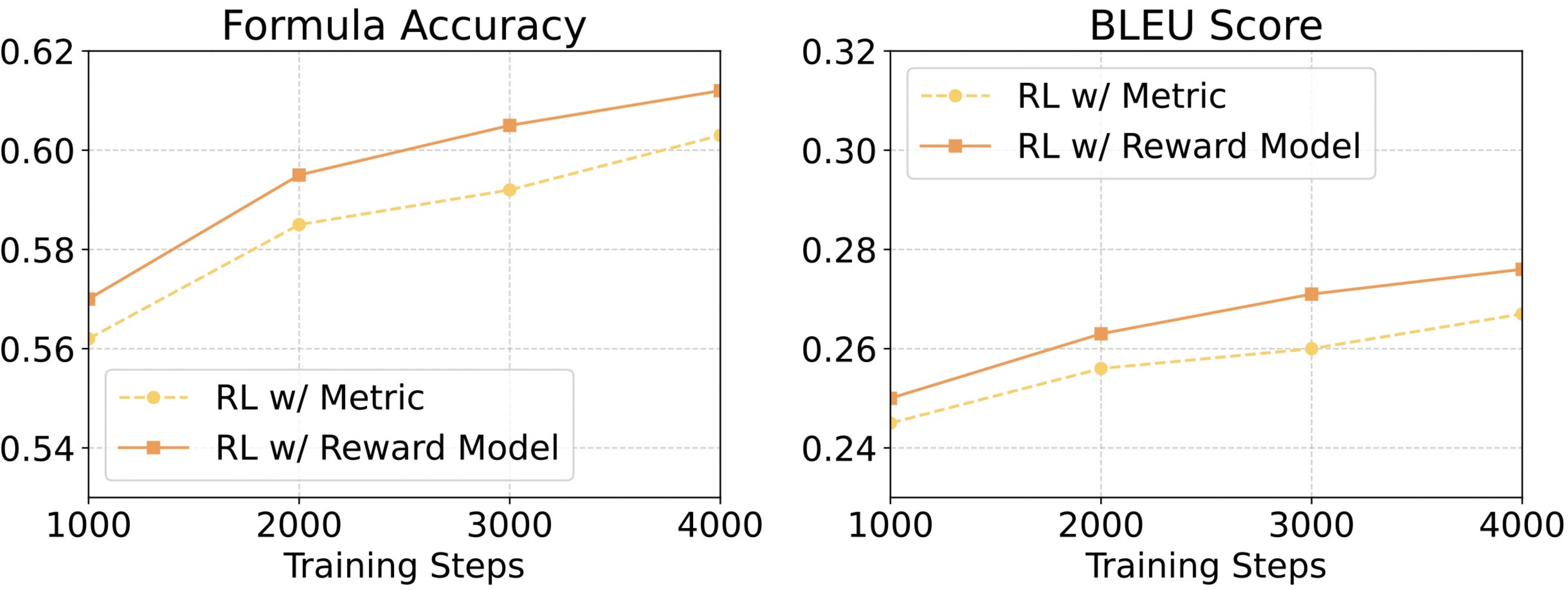}
    \caption{Comparison of reinforcement learning feedback strategies: reward model vs. metric supervision on STL-DivEn.}
    \label{fig:training_curve}
\end{figure}

To compare reward model feedback with direct metric supervision in reinforcement learning, we evaluate their performance on the STL-DivEn dataset. As shown in Figure~\ref{fig:training_curve}, models trained with reward models achieve higher gains in formula and template accuracy. Specifically, template accuracy reaches 61.2\% with reward models versus 60.3\% with metric-based supervision, and BLEU improves from 0.267 to 0.276. This advantage is due to the reward model’s ability to capture deeper NL-STL correspondence through preference learning, while metric supervision may introduce noise. These results suggest reward models provide more effective guidance for RL in this task.

\section{Conclusion}
%This section summarizes the key findings of the research and discusses its implications. It also suggests potential future work and directions for further research.
In this work, we propose RESTL, a reinforcement learning framework for transforming natural language into STL specifications. It employs multi-aspect reward models to ensure semantic correctness and uses curriculum learning to improve training efficiency. Experiments on two benchmarks show that RESTL outperforms existing methods in formula accuracy, template accuracy, and BLEU. By integrating structured rewards with progressive training, RESTL provides an effective solution for formal specification generation of cyber-physical systems.

% In this work, we present RESTL, a reinforcement learning-based framework for transforming natural language descriptions into STL specifications. 
% RESTL introduces a multi-aspect reward modeling strategy that captures semantic correctness through dedicated reward models. We further incorporate curriculum learning to enhance the efficiency of the reward model.
% Comprehensive experiments on two benchmark datasets indicate that RESTL significantly outperforms existing methods in formula accuracy, template accuracy, and BLEU. 
% RESTL offers a flexible and extensible paradigm for formal language generation. By combining reinforcement learning with structured reward signals and progressive training, it bridges the gap between natural language expressiveness and formal specification rigor. 
% In future work, we plan to extend RESTL to integrate symbolic verification tools. 
% \clearpage
\section*{Acknowledgement}
We sincerely thank the anonymous reviewers for their valuable comments and suggestions. This work is supported by the National Natural Science Foundation of China under Grant Nos. 62192731, 62192732, 62192730, and 62272166.
\bibliography{aaai2026}

\begin{thebibliography}{39}
\providecommand{\natexlab}[1]{#1}

\bibitem[{Bengio et~al.(2009)Bengio, Louradour, Collobert, and Weston}]{bengio2009curriculum}
Bengio, Y.; Louradour, J.; Collobert, R.; and Weston, J. 2009.
\newblock Curriculum learning.
\newblock In \emph{Proceedings of the 26th annual international conference on machine learning}, 41--48.

\bibitem[{Bradley and Terry(1952)}]{bradley1952rank}
Bradley, R.~A.; and Terry, M.~E. 1952.
\newblock Rank analysis of incomplete block designs: I. The method of paired comparisons.
\newblock \emph{Biometrika}, 39(3/4): 324--345.

\bibitem[{Chen et~al.(2023)Chen, Gandhi, Zhang, and Fan}]{ChenGZF23}
Chen, Y.; Gandhi, R.; Zhang, Y.; and Fan, C. 2023.
\newblock {NL2TL:} Transforming Natural Languages to Temporal Logics using Large Language Models.
\newblock In Bouamor, H.; Pino, J.; and Bali, K., eds., \emph{Proceedings of the 2023 Conference on Empirical Methods in Natural Language Processing, {EMNLP} 2023, Singapore, December 6-10, 2023}, 15880--15903. Association for Computational Linguistics.

\bibitem[{Christiano et~al.(2017)Christiano, Leike, Brown, Martic, Legg, and Amodei}]{christiano2017deep}
Christiano, P.~F.; Leike, J.; Brown, T.; Martic, M.; Legg, S.; and Amodei, D. 2017.
\newblock Deep reinforcement learning from human preferences.
\newblock \emph{Advances in neural information processing systems}, 30.

\bibitem[{Cosler et~al.(2023)Cosler, Hahn, Mendoza, Schmitt, and Trippel}]{CoslerHMST23}
Cosler, M.; Hahn, C.; Mendoza, D.; Schmitt, F.; and Trippel, C. 2023.
\newblock nl2spec: Interactively Translating Unstructured Natural Language to Temporal Logics with Large Language Models.
\newblock In Enea, C.; and Lal, A., eds., \emph{Computer Aided Verification - 35th International Conference, {CAV} 2023, Paris, France, July 17-22, 2023, Proceedings, Part {II}}, volume 13965 of \emph{Lecture Notes in Computer Science}, 383--396. Springer.

\bibitem[{Dann, Mansour, and Mohri(2023)}]{dann2023reinforcement}
Dann, C.; Mansour, Y.; and Mohri, M. 2023.
\newblock Reinforcement learning can be more efficient with multiple rewards.
\newblock In \emph{International Conference on Machine Learning}, 6948--6967. PMLR.

\bibitem[{Dwyer, Avrunin, and Corbett(1999)}]{dwyer1999patterns}
Dwyer, M.~B.; Avrunin, G.~S.; and Corbett, J.~C. 1999.
\newblock Patterns in property specifications for finite-state verification.
\newblock In \emph{Proceedings of the 21st international conference on Software engineering}, 411--420.

\bibitem[{Ernst et~al.(2022)Ernst, Arcaini, Fainekos, Formica, Inoue, Khandait, Mahboob, Menghi, Pedrielli, Waga, Yamagata, and Zhang}]{ARCHCOMP22Falsification}
Ernst, G.; Arcaini, P.; Fainekos, G.; Formica, F.; Inoue, J.; Khandait, T.; Mahboob, M.~M.; Menghi, C.; Pedrielli, G.; Waga, M.; Yamagata, Y.; and Zhang, Z. 2022.
\newblock {ARCH-COMP} 2022 Category Report: Falsification with Ubounded Resources.
\newblock In Frehse, G.; Althoff, M.; Schoitsch, E.; and Guiochet, J., eds., \emph{Proceedings of 9th International Workshop on Applied Verification of Continuous and Hybrid Systems (ARCH22)}, volume~90 of \emph{EPiC Series in Computing}, 204--221. EasyChair.

\bibitem[{Fang et~al.(2025)Fang, Jin, An, Chen, Chen, and Zhan}]{fang2025enhancingtransformationnaturallanguage}
Fang, Y.; Jin, Z.; An, J.; Chen, H.; Chen, X.; and Zhan, N. 2025.
\newblock Enhancing Transformation from Natural Language to Signal Temporal Logic Using LLMs with Diverse External Knowledge.
\newblock In Che, W.; Nabende, J.; Shutova, E.; and Pilehvar, M.~T., eds., \emph{Findings of the Association for Computational Linguistics, {ACL} 2025, Vienna, Austria, July 27 - August 1, 2025}, 10446--10458. Association for Computational Linguistics.

\bibitem[{Ghosh et~al.(2016)Ghosh, Elenius, Li, Lincoln, Shankar, and Steiner}]{ghosh2016arsenal}
Ghosh, S.; Elenius, D.; Li, W.; Lincoln, P.; Shankar, N.; and Steiner, W. 2016.
\newblock ARSENAL: automatic requirements specification extraction from natural language.
\newblock In \emph{NASA Formal Methods: 8th International Symposium, NFM 2016, Minneapolis, MN, USA, June 7-9, 2016, Proceedings 8}, 41--46. Springer.

\bibitem[{Graves et~al.(2017)Graves, Bellemare, Menick, Munos, and Kavukcuoglu}]{graves2017automated}
Graves, A.; Bellemare, M.~G.; Menick, J.; Munos, R.; and Kavukcuoglu, K. 2017.
\newblock Automated curriculum learning for neural networks.
\newblock In \emph{international conference on machine learning}, 1311--1320. Pmlr.

\bibitem[{Guo et~al.(2025)Guo, Yang, Zhang, Song, Zhang, Xu, Zhu, Ma, Wang, Bi et~al.}]{guo2025deepseek}
Guo, D.; Yang, D.; Zhang, H.; Song, J.; Zhang, R.; Xu, R.; Zhu, Q.; Ma, S.; Wang, P.; Bi, X.; et~al. 2025.
\newblock Deepseek-r1: Incentivizing reasoning capability in llms via reinforcement learning.
\newblock \emph{arXiv preprint arXiv:2501.12948}.

\bibitem[{He et~al.(2022)He, Bartocci, Nickovic, Isakovic, and Grosu}]{HeBNIG22}
He, J.; Bartocci, E.; Nickovic, D.; Isakovic, H.; and Grosu, R. 2022.
\newblock DeepSTL - From English Requirements to Signal Temporal Logic.
\newblock In \emph{44th {IEEE/ACM} 44th International Conference on Software Engineering, {ICSE} 2022, Pittsburgh, PA, USA, May 25-27, 2022}, 610--622. {ACM}.

\bibitem[{Justesen et~al.(2018)Justesen, Rodriguez~Torrado, Bontrager, Khalifa, Togelius, and Risi}]{justesen2018illuminating}
Justesen, N.; Rodriguez~Torrado, R.; Bontrager, P.; Khalifa, A.; Togelius, J.; and Risi, S. 2018.
\newblock Illuminating Generalization in Deep Reinforcement Learning through Procedural Level Generation.
\newblock In \emph{NeurIPS Workshop on Deep Reinforcement Learning}.

\bibitem[{Kingma(2014)}]{kingma2014adam}
Kingma, D.~P. 2014.
\newblock Adam: A method for stochastic optimization.
\newblock \emph{arXiv preprint arXiv:1412.6980}.

\bibitem[{Kulkarni, Fisher, and Myers(2013)}]{kulkarni2013new}
Kulkarni, D.; Fisher, A.~N.; and Myers, C.~J. 2013.
\newblock A new assertion property language for analog/mixed-signal circuits.
\newblock In \emph{Proceedings of the 2013 Forum on specification and Design Languages (FDL)}, 1--8. IEEE.

\bibitem[{Li et~al.(2020)Li, Jabri, Darrell, and Agrawal}]{li2020towards}
Li, R.; Jabri, A.; Darrell, T.; and Agrawal, P. 2020.
\newblock Towards practical multi-object manipulation using relational reinforcement learning.
\newblock In \emph{2020 ieee international conference on robotics and automation (icra)}, 4051--4058. IEEE.

\bibitem[{Lignos et~al.(2015)Lignos, Raman, Finucane, Marcus, and Kress{-}Gazit}]{LignosRFMK15}
Lignos, C.; Raman, V.; Finucane, C.; Marcus, M.~P.; and Kress{-}Gazit, H. 2015.
\newblock Provably correct reactive control from natural language.
\newblock \emph{Auton. Robots}, 38(1): 89--105.

\bibitem[{Lin(2004)}]{lin2004rouge}
Lin, C.-Y. 2004.
\newblock Rouge: A package for automatic evaluation of summaries.
\newblock In \emph{Text summarization branches out}, 74--81.

\bibitem[{Madsen et~al.(2018)Madsen, Vaidyanathan, Sadraddini, Vasile, DeLateur, Weiss, Densmore, and Belta}]{MadsenVSVDWDB18}
Madsen, C.; Vaidyanathan, P.; Sadraddini, S.; Vasile, C.~I.; DeLateur, N.~A.; Weiss, R.; Densmore, D.; and Belta, C. 2018.
\newblock Metrics for Signal Temporal Logic Formulae.
\newblock In \emph{57th {IEEE} Conference on Decision and Control, {CDC} 2018, Miami, FL, USA, December 17-19, 2018}, 1542--1547. {IEEE}.

\bibitem[{Maierhofer et~al.(2020)Maierhofer, Rettinger, Mayer, and Althoff}]{maierhofer2020formalization}
Maierhofer, S.; Rettinger, A.-K.; Mayer, E.~C.; and Althoff, M. 2020.
\newblock Formalization of interstate traffic rules in temporal logic.
\newblock In \emph{2020 IEEE Intelligent Vehicles Symposium (IV)}, 752--759. IEEE.

\bibitem[{Maler and Ni{\v{c}}kovi{\'c}(2004)}]{maler2004monitoring}
Maler, O.; and Ni{\v{c}}kovi{\'c}, D. 2004.
\newblock Monitoring temporal properties of continuous signals.
\newblock In \emph{{FORMATS/FTRTFT} 2004}, volume 3253 of \emph{LNCS}, 152--166. Springer.

\bibitem[{Mnih et~al.(2016)Mnih, Badia, Mirza, Graves, Lillicrap, Harley, Silver, and Kavukcuoglu}]{mnih2016asynchronous}
Mnih, V.; Badia, A.~P.; Mirza, M.; Graves, A.; Lillicrap, T.; Harley, T.; Silver, D.; and Kavukcuoglu, K. 2016.
\newblock Asynchronous methods for deep reinforcement learning.
\newblock In \emph{International conference on machine learning}, 1928--1937. PmLR.

\bibitem[{Papineni et~al.(2002)Papineni, Roukos, Ward, and Zhu}]{papineni2002bleu}
Papineni, K.; Roukos, S.; Ward, T.; and Zhu, W.-J. 2002.
\newblock Bleu: a method for automatic evaluation of machine translation.
\newblock In \emph{Proceedings of the 40th annual meeting of the Association for Computational Linguistics}, 311--318.

\bibitem[{Paszke(2019)}]{paszke2019pytorch}
Paszke, A. 2019.
\newblock Pytorch: An imperative style, high-performance deep learning library.
\newblock \emph{arXiv preprint arXiv:1912.01703}.

\bibitem[{Pnueli(1977)}]{pnueli1977temporal}
Pnueli, A. 1977.
\newblock The Temporal Logic of Programs.
\newblock In \emph{{FOCS} 1977}, 46--57. {IEEE}.

\bibitem[{Ryu et~al.(2024)Ryu, Do, Kim, Lee, and Ok}]{ryu2024multi}
Ryu, S.; Do, H.; Kim, Y.; Lee, G.; and Ok, J. 2024.
\newblock Multi-Dimensional Optimization for Text Summarization via Reinforcement Learning.
\newblock In \emph{Proceedings of the 62nd Annual Meeting of the Association for Computational Linguistics (Volume 1: Long Papers)}, 5858--5871.

\bibitem[{Santos, Carvalho, and Sampaio(2018)}]{santos2018formal}
Santos, T.; Carvalho, G.; and Sampaio, A. 2018.
\newblock Formal modelling of environment restrictions from natural-language requirements.
\newblock In \emph{Formal Methods: Foundations and Applications: 21st Brazilian Symposium, SBMF 2018, Salvador, Brazil, November 26--30, 2018, Proceedings 21}, 252--270. Springer.

\bibitem[{Schulman et~al.(2017)Schulman, Wolski, Dhariwal, Radford, and Klimov}]{schulman2017proximal}
Schulman, J.; Wolski, F.; Dhariwal, P.; Radford, A.; and Klimov, O. 2017.
\newblock Proximal policy optimization algorithms.
\newblock \emph{arXiv preprint arXiv:1707.06347}.

\bibitem[{Team et~al.(2025)Team, Du, Gao, Xing, Jiang, Chen, Li, Xiao, Du, Liao et~al.}]{team2025kimi}
Team, K.; Du, A.; Gao, B.; Xing, B.; Jiang, C.; Chen, C.; Li, C.; Xiao, C.; Du, C.; Liao, C.; et~al. 2025.
\newblock Kimi k1. 5: Scaling reinforcement learning with llms.
\newblock \emph{arXiv preprint arXiv:2501.12599}.

\bibitem[{Tellex et~al.(2020)Tellex, Gopalan, Kress-Gazit, and Matuszek}]{tellex2020robots}
Tellex, S.; Gopalan, N.; Kress-Gazit, H.; and Matuszek, C. 2020.
\newblock Robots that use language.
\newblock \emph{Annual Review of Control, Robotics, and Autonomous Systems}, 3(1): 25--55.

\bibitem[{Wang et~al.(2019)Wang, Lehman, Clune, and Stanley}]{wang2019paired}
Wang, R.; Lehman, J.; Clune, J.; and Stanley, K.~O. 2019.
\newblock Paired open-ended trailblazer (poet): Endlessly generating increasingly complex and diverse learning environments and their solutions.
\newblock \emph{arXiv preprint arXiv:1901.01753}.

\bibitem[{Wang et~al.(2025)Wang, Zhang, Pang, Guo, Zheng, and Zheng}]{wang2025maferw}
Wang, Y.; Zhang, H.; Pang, L.; Guo, B.; Zheng, H.; and Zheng, Z. 2025.
\newblock MaFeRw: Query rewriting with multi-aspect feedbacks for retrieval-augmented large language models.
\newblock In \emph{Proceedings of the AAAI Conference on Artificial Intelligence}, volume~39, 25434--25442.

\bibitem[{Wolf et~al.(2020)Wolf, Debut, Sanh, Chaumond, Delangue, Moi, Cistac, Rault, Louf, Funtowicz et~al.}]{wolf2020transformers}
Wolf, T.; Debut, L.; Sanh, V.; Chaumond, J.; Delangue, C.; Moi, A.; Cistac, P.; Rault, T.; Louf, R.; Funtowicz, M.; et~al. 2020.
\newblock Transformers: State-of-the-art natural language processing.
\newblock In \emph{Proceedings of the 2020 conference on empirical methods in natural language processing: system demonstrations}, 38--45.

\bibitem[{Xie et~al.(2025)Xie, Gao, Ren, Luo, Hong, Dai, Zhou, Qiu, Wu, and Luo}]{xie2025logic}
Xie, T.; Gao, Z.; Ren, Q.; Luo, H.; Hong, Y.; Dai, B.; Zhou, J.; Qiu, K.; Wu, Z.; and Luo, C. 2025.
\newblock Logic-rl: Unleashing llm reasoning with rule-based reinforcement learning.
\newblock \emph{arXiv preprint arXiv:2502.14768}.

\bibitem[{Zheng et~al.(2023)Zheng, Dou, Gao, Hua, Shen, Wang, Liu, Jin, Liu, Zhou et~al.}]{zheng2023secrets}
Zheng, R.; Dou, S.; Gao, S.; Hua, Y.; Shen, W.; Wang, B.; Liu, Y.; Jin, S.; Liu, Q.; Zhou, Y.; et~al. 2023.
\newblock Secrets of rlhf in large language models part i: Ppo.
\newblock \emph{arXiv preprint arXiv:2307.04964}.

\bibitem[{Zheng et~al.(2024)Zheng, Zhang, Zhang, YeYanhan, and Luo}]{zheng2024llamafactory}
Zheng, Y.; Zhang, R.; Zhang, J.; YeYanhan, Y.; and Luo, Z. 2024.
\newblock LlamaFactory: Unified Efficient Fine-Tuning of 100+ Language Models.
\newblock In \emph{Proceedings of the 62nd Annual Meeting of the Association for Computational Linguistics (Volume 3: System Demonstrations)}, 400--410.

\bibitem[{Ziegler et~al.(2019)Ziegler, Stiennon, Wu, Brown, Radford, Amodei, Christiano, and Irving}]{ziegler2019fine}
Ziegler, D.~M.; Stiennon, N.; Wu, J.; Brown, T.~B.; Radford, A.; Amodei, D.; Christiano, P.; and Irving, G. 2019.
\newblock Fine-tuning language models from human preferences.
\newblock \emph{arXiv preprint arXiv:1909.08593}.

\bibitem[{{\v{Z}}ilka(2010)}]{vzilka2010temporal}
{\v{Z}}ilka, L. 2010.
\newblock \emph{Temporal logic for man}.
\newblock Ph.D. thesis, Master’s thesis, Brno University of Technology.

\end{thebibliography}

\clearpage
\clearpage
\appendix
\section{A \; The Appendix of Related Work}

\subsection{Reinforcement Learning for LLMs}

% With the rapid development of large language models (LLMs), reinforcement learning (RL) has been widely adopted to further improve their capabilities.
% One representative approach is reinforcement learning from human feedback (RLHF), which optimizes model performance using reward scores generated by reward models, combined with policy gradient algorithms such as Proximal Policy Optimization (PPO).
% To provide more fine-grained supervision signals, prior work has introduced critic models to compute intermediate-stage rewards.
% Recent studies have shown that reinforcement learning with multiple reward signals can lead to more efficient training.
% In the field of summarization, Su et al. (2023) proposed a multi-reward reinforcement learning framework for multi-document summarization, constructing separate policy models for importance, redundancy, and length. By introducing two multi-dimensional optimization strategies (MDOmin and MDOpro), their approach adaptively optimizes summary quality across four dimensions: consistency, coherence, relevance, and fluency.
% In the RAG domain, MaFeRw introduces a novel query rewriting method that integrates multi-aspect feedback from both retrieval and generation, using reinforcement learning with multiple rewards to significantly improve generation quality and training stability within RAG systems.
% Inspired by these works, in this study, we apply reinforcement learning with multi-aspect dense reward signals to enhance the model’s ability to transform natural language into accurate STL specifications.

With the rapid development of LLMs, reinforcement learning (RL) has been widely adopted to further improve their capabilities~\cite{ziegler2019fine,christiano2017deep}.
One representative approach is reinforcement learning from human feedback (RLHF)~\cite{christiano2017deep}, which optimizes model performance using reward scores generated by reward models, combined with policy gradient algorithms such as PPO~\cite{schulman2017proximal}.
% Due to the inherent instability of RL training procedures, recent studies have adopted supervised fine-tuning (SFT) to approximate the effects of reinforcement learning~\cite{rafailov2023direct, liu2024chain, lu2022quark, zhao2023slic}. These methods integrate response quality directly into the supervision signals, leading to more stable and effective training. 
To provide more fine-grained supervision signals, prior work has introduced critic models to compute intermediate-stage rewards~\cite{mnih2016asynchronous, zheng2023secrets}.
Recent studies have shown that reinforcement learning with multiple reward signals can lead to more efficient training~\cite{dann2023reinforcement}.
%~\cite{liu2023chain, rafailov2023direct}.
For example, multi-dimensional optimization has been used to balance various quality aspects such as consistency and relevance~\cite{ryu2024multi}. In the RAG domain, MaFeRw~\cite{wang2025maferw} integrates multi-aspect feedback from both retrieval and generation stages to improve output quality and training stability.
In this study, we apply reinforcement learning with multi-aspect dense reward signals to enhance the model’s ability to transform natural language descriptions into accurate STL specifications.

\subsection{Curriculum Learning for LLMs}

Curriculum Learning (CL)\cite{bengio2009curriculum,graves2017automated} is a training paradigm that organizes training samples in order, guiding the model to learn from easier examples before gradually progressing to more complex ones. 
In the context of Reinforcement Learning (RL), curricula are typically designed for specific tasks by incrementally increasing the complexity of the environment\cite{justesen2018illuminating,wang2019paired,li2020towards}, thereby improving the generalization and transferability of the learned policies. 
% To reduce reliance on manually crafted task hierarchies, researchers have proposed teacher-guided curriculum learning~\cite{matiisen2019teacher,portelas2020teacher}, where a teacher agent dynamically selects tasks for a student agent based on its learning state. This task selection process is often modeled as a Partially Observable Markov Decision Process (POMDP).
Recently, with the growing application of RL in the post-training phase of LLMs, curriculum learning has shown great potential for improving both training efficiency and model performance. 
% Among existing approaches, some methods employ manually designed heuristic curricula with fixed training stages. 
For instance, Kimi k1.5~\cite{team2025kimi} and LogicRL~\cite{xie2025logic} train models on ``easy'' samples for a fixed number of steps before switching to more ``difficult'' examples.
In this work, we apply curriculum learning to train the reward models by ranking NL-to-STL transformation difficulty across multiple dimensions, enabling them to start from simple samples and progressively adapt to accurately evaluate more difficult tasks.
\section{B \; The Appendix of Method}

\subsection{Details of Reinforcement Learning in RESTL}

\begin{algorithm}[t]
\caption{Training STL Generator with Multi-Reward PPO}
\label{alg:stl-ppo}
\KwIn{
Training data \(D = \{x^{(i)}\}_{i=1}^N\), pretrained generator \(G_{\theta_0}\), reward models \(r_a, r_t, r_l, r_s\), reward weights \(\lambda_1, \lambda_2, \lambda_3, \lambda_4\), PPO parameters, KL coefficient \(\eta\)
}
\KwOut{Optimized STL generator \(G_{\theta}\)}

Initialize generator \(G_{\theta} \gets G_{\theta_0}\)\;

\For{each training epoch}{ \label{line:each_epoch}
    \ForEach{\(x^{(i)} \in D\)}{ \label{line:each_sample}
        \tcp{Sample a generated STL formula}
        Sample \(\hat{y}^{(i)} \sim G_{\theta}(x^{(i)})\)\; \label{line:sample_formula}

        \tcp{Compute multi-aspect rewards}
        Compute atomic proposition alignment: \(r_a^{(i)} \gets r_a(x^{(i)}, \hat{y}^{(i)})\)\;\label{line:reward_start}
        
        Compute templated NL similarity: \(r_t^{(i)} \gets r_t(x^{(i)}, \hat{y}^{(i)})\)\;
        
        Compute formula succinctness: \(r_l^{(i)} \gets r_l(x^{(i)}, \hat{y}^{(i)})\)\;
        
        Compute STL-level similarity: \(r_s^{(i)} \gets r_s(x^{(i)}, \hat{y}^{(i)})\)\;

        \tcp{Aggregate overall reward}
        \(r_{\text{RL}}^{(i)} \gets \lambda_1 r_a^{(i)} + \lambda_2 r_t^{(i)} + \lambda_3 r_l^{(i)} + \lambda_4 r_s^{(i)}\)\;

        \tcp{Apply KL-regularized objective}
        \(r_{\text{total}}^{(i)} \gets r_{\text{RL}}^{(i)} - \eta \cdot \text{KL}(G_{\theta} \parallel G_{\theta_0})\)\; \label{line:reward_end}

        \tcp{Update generator using PPO}
        Compute PPO loss: \(\mathcal{L} \gets \text{PPO\_Loss}(r_{\text{total}}^{(i)}, G_{\theta})\)\; \label{line:apply_ppo_start}

        Update policy parameters: \(\theta \gets \theta - \nabla_\theta \mathcal{L}\)\; \label{line:apply_ppo_end}
    }
}
\Return \(G_{\theta}\)\;
\end{algorithm}

The RL training process of RESTL is shown in Algorithm~\ref{alg:stl-ppo}. It starts with the prepared initial STL generator \( G_{\theta_0} \).

%and iteratively updates the policy using PPO. For each training example, the generator samples multiple candidate STL formulas. Each candidate is evaluated with four reward models. These rewards are aggregated and used to compute the PPO loss with a KL penalty. Finally, the generator is updated using gradient descent.

In each training epoch, we compute the aggregated multi-aspect reward for each sample $x^{(i)} \in D$ (Line~\ref{line:each_sample}). First, for each NL description $x^{(i)}$, we generate a corresponding STL formula $\hat{y}^{(i)}$ (Line~\ref{line:sample_formula}). Then, we compute its scores using the four reward models and aggregate them into an overall reward (Line~\ref{line:reward_start} to \ref{line:reward_end}). Formally, given a natural language instruction \( x \) and the generated STL formula \( \hat{y} \), the overall reward is computed as:
\begin{equation*}
r_{\text{RL}}(\hat{y}) = \lambda_1 r_{\text{a}}(\hat{y}) + \lambda_2 r_{\text{t}}(\hat{y}) + \lambda_3 r_{\text{l}}(\hat{y}) + \lambda_4 r_{\text{s}}(\hat{y}),
% \label{eq:stl-reward}
\end{equation*}
where \( r_{\text{a}} \), \( r_{\text{t}} \), \( r_{\text{l}} \), and \( r_{\text{s}} \) are the reward scores corresponding to the atomic proposition alignment $m_{\text{a}}$, templated NL similarity $m_{\text{t}}$, formula conciseness $m_{\text{l}}$, and STL-level similarity $m_{\text{s}}$, proposed in Section~\ref{subsec:feedbacks}, respectively. The hyperparameters \( \lambda_1, \lambda_2, \lambda_3, \lambda_4 \) control the relative importance of each reward. 
The reward objective $r_{\text{total}}$ is to maximize expected reward while constraining deviation from the initial policy \( G_{\theta_0} \). 
%The objective is given by:
\begin{equation*}
r_{\text{total}} = r_{\text{RL}}(\hat{y}) - \eta \cdot \text{KL}(G_{\theta} \parallel G_{\theta_0}),
% \label{eq:ppo-loss}
\end{equation*}
where \( \eta \) is a KL penalty coefficient. This term stabilizes training by penalizing large shifts from the pre-trained generator.

% Finally, we apply the Proximal Policy Optimization (PPO) algorithm~\cite{schulman2017proximal} to optimize the STL generator \( G_{\theta} \) (Line~\ref{line:apply_ppo_start} to \ref{line:apply_ppo_end}). 
Finally, we apply the PPO algorithm~\cite{schulman2017proximal} to optimize the STL generator \( G_{\theta} \) using the KL-regularized reward signal (Line~\ref{line:apply_ppo_start} to \ref{line:apply_ppo_end}).

\begin{table*}[t]
    \centering
    \resizebox{\textwidth}{!}{
    \begin{tabular}{lcccccc}
        \toprule
        \multirow{2}{*}{\textbf{Model}} 
        & \multicolumn{3}{c}{\textbf{STL-DivEn}} 
        & \multicolumn{3}{c}{\textbf{DeepSTL}} \\
        \cmidrule(lr){2-4} \cmidrule(lr){5-7}
        & \textbf{Formula Acc.} & \textbf{Template Acc.} & \textbf{BLEU} 
        & \textbf{Formula Acc.} & \textbf{Template Acc.} & \textbf{BLEU} \\
        \midrule
        DeepSTL & 0.1986 $\pm$ 0.0182 & 0.1883 $\pm$ 0.0109 & 0.0293 $\pm$ 0.0043 
                & 0.2002 $\pm$ 0.0146 & 0.2916 $\pm$ 0.0172 & 0.3332 $\pm$ 0.0192 \\
        GPT-3.5 & 0.3018 $\pm$ 0.0208 & 0.3034 $\pm$ 0.0223 & 0.0424 $\pm$ 0.0182 
                & 0.2145 $\pm$ 0.0137 & 0.3002 $\pm$ 0.0162 & 0.2249 $\pm$ 0.0203 \\
        GPT-4   & 0.4733 $\pm$ 0.0284 & 0.4741 $\pm$ 0.0293 & 0.0831 $\pm$ 0.0196 
                & 0.2262 $\pm$ 0.0172 & 0.3048 $\pm$ 0.0192 & 0.2881 $\pm$ 0.0217 \\
        DeepSeek & 0.4790 $\pm$ 0.0195 & 0.4825 $\pm$ 0.0206 & 0.0791 $\pm$ 0.0175 
                 & 0.2537 $\pm$ 0.0182 & 0.3254 $\pm$ 0.0193 & 0.3982 $\pm$ 0.0233 \\
        KGST    & 0.5587 $\pm$ 0.0263 & 0.5627 $\pm$ 0.0276 & 0.2142 $\pm$ 0.0187 
                & 0.4538 $\pm$ 0.0207 & 0.4939 $\pm$ 0.0264 & 0.5686 $\pm$ 0.0301 \\
        \textbf{RESTL (Ours)} & \textbf{0.6838 $\pm$ 0.0243} & \textbf{0.6974 $\pm$ 0.0255} & \textbf{0.3347 $\pm$ 0.0177} 
                              & \textbf{0.5985 $\pm$ 0.0183} & \textbf{0.6327 $\pm$ 0.0192} & \textbf{0.6783 $\pm$ 0.0177} \\
        \bottomrule
    \end{tabular}
    }
    \caption{Experimental results on performance and variability measures of RESTL and baselines on STL-DivEn and DeepSTL datasets.}
    \label{tab:full_main_result}
\end{table*}

\section{C \; The Appendix of Evaluation Metric}

% Let $y^* = [y_1^*, y_2^*, \ldots, y_n^*]$ be the reference token sequence and $\hat{y} = [\hat{y}_1, \hat{y}_2, \ldots, \hat{y}_n]$ be the predicted sequence of the same length. The formula accuracy is computed as:
% \[
% A_F = \frac{1}{n} \sum_{i=1}^{n} \mathbb{I}[y_i^* = \hat{y}_i],
% \]
% which, \(\mathbb{I}[\cdot]\) denotes the \textbf{indicator function}, which evaluates whether the condition inside the brackets holds. If the condition is true, then \(\mathbb{I}[\cdot] = 1\); otherwise, it equals 0.
\paragraph{STL Formula Accuracy} 
Measure the alignment accuracy between the reference and prediction sequences at the string level.

\paragraph{Template Accuracy} 
First transform the reference and prediction instances into STL templates, then calculate the alignment accuracy of the resulting template sequences.
% Let $\text{Tem}(y^*)$ and $\text{Tem}(\hat{y})$ denote the abstracted STL templates of the reference and predicted formulas, respectively, each consisting of $m$ tokens. The template accuracy is defined similarly:
% \[
% A_\text{T} = \frac{1}{m} \sum_{j=1}^{m} \mathbb{I}[\text{Tem}(y_j^*) = \text{Tem}(\hat{y}_j)].
% \]

The following example shows both the formula and its corresponding template for the reference and predicted outputs.
\[
\scalebox{0.85}{$
\begin{aligned}
\text{Formula:} \; & \G ( x > 8 ) \rightarrow \F ( y < 3 )
\Rightarrow \, \text{Template:} \;  \G ( \phi ) \rightarrow \F ( \phi ) \\
\text{Formula:} \; & \G ( x > 8 ) \rightarrow \F ( z < 3 )  
\Rightarrow \, \text{Template:} \;  \G ( \phi ) \rightarrow \F ( \phi )
\end{aligned}
$}
\]
The formula contains 13 tokens. All tokens coincide except for the variable ``y'' vs ``z'' in the atomic proposition, resulting in $A_F = \frac{12}{13}$. When both formulas are converted to templates by replacing atomic propositions with placeholders (e.g., $\phi$), all tokens align perfectly: $A_\text{T} = 1$.

\section{D \; The Appendix of Implementation Details}
Our experiments are carried out on 8 NVIDIA GeForce RTX 4090 GPUs (24GB VRAM each). We implement our framework using PyTorch~\cite{paszke2019pytorch} and Huggingface Transformers~\cite{wolf2020transformers}, with LLaMA-Factory~\cite{zheng2024llamafactory} as the base for model customization.
Each reward model is fine-tuned in LLaMA 3-8B with a linear value head and trained for 5 epochs using Adam optimizer~\cite{kingma2014adam}, with a learning rate of 5e-5 and a batch size of 16. 
The STL generator is fine-tuned using PPO for 80,000 steps, with a batch size of 32, a learning rate of 1.41e-5, and a KL penalty coefficient $\eta$ set to 0.05. The combined reward signal is computed using weighted scores with $\lambda_1 = 0.2$, $\lambda_2 = 0.25$, $\lambda_3 = 0.35$, and $\lambda_4 = 0.2$, where these hyperparameters control the weights of different reward components. Their initial values were assigned based on the proportional contribution of each individual metric to the model’s performance improvement. Subsequently, grid search was performed to fine-tune these weights, resulting in the final values above. We found that this combination achieves superior performance across different datasets.

\section{E \; The Appendix of Experimental Results}

\subsection{E.1 Experimental Results on Variability Measures}

Table~\ref{tab:full_main_result} presents experimental results on performance and variability measures of RESTL and baseline models on the STL-DivEn and DeepSTL datasets. It is evident from the table that RESTL outperforms all other baselines across both datasets.

\subsection{E.2 Ablation Study Details}

\begin{table}[t]
\centering
% \caption{Performance comparison using different reward signals for RL training on STL-DivEn and DeepSTL datasets.}
\begin{subtable}[t]{1\linewidth}
    \centering
    \caption{STL-DivEn dataset}
    \resizebox{\linewidth}{!}{
    \begin{tabular}{lcccc}
    \toprule
    \textbf{Model} & \textbf{Formula Acc.} & \textbf{Template Acc.} & \textbf{BLEU}\\
    \midrule
    RESTL      & \textbf{0.6838} & \textbf{0.6974} & \textbf{0.3347}\\  
    $m_{\text{a}}$    & 0.6027 & 0.6152 & 0.2824\\
    $m_{\text{t}}$       & 0.6172 & 0.6137 & 0.2652 \\
    $m_{\text{l}}$       & 0.5952 & 0.6072 & 0.2636  \\
    $m_{\text{s}}$       & 0.6282 & 0.6299 & 0.2892   \\
    \midrule
    LLaMA3 (Fine-tuned) & 0.4956 & 0.5007 & 0.1784 \\
    \bottomrule
    \end{tabular}
    }
\end{subtable}
% \vspace{2mm}

\begin{subtable}[t]{1\linewidth}
    \centering
    \caption{DeepSTL dataset}
    \resizebox{\linewidth}{!}{
    \begin{tabular}{lccc}
    \toprule
    \textbf{Model} & \textbf{Formula Acc.} & \textbf{Template Acc.} & \textbf{BLEU} \\
    \midrule
    RESTL      & \textbf{0.5985} & \textbf{0.6327} & \textbf{0.6783}\\
    $m_{\text{a}}$   & 0.5203 & 0.5182 & 0.5872 \\
    $m_{\text{t}}$       & 0.5182 & 0.5191 & 0.5925 \\
    $m_{\text{l}}$       & 0.4952 & 0.5001 & 0.5782  \\
    $m_{\text{s}}$       & 0.5373 & 0.5407 & 0.6093 \\
    \midrule
    LLaMA3 (Fine-tuned) & 0.2850 & 0.3285 & 0.5579 \\
    \bottomrule
    \end{tabular}
    }
\end{subtable}
\caption{Ablation analysis of different reward feedback mechanisms on each metric for STL-DivEn and DeepSTL datasets.}
\label{tab:reward_rl_individual}
\end{table}

From Table~\ref{tab:reward_rl_individual}, we can see that each individual reward metric contributes positively to model performance, outperforming LLaMA3 (Fine-tuned) but falling short of RESTL.

% From Table~\ref{tab:reward_rl_individual}, we can see that each individual reward metric is effective, with performance higher than LLaMA3 (Fine-tuned) but lower than RESTL.

% Reward metric $m_a$ achieves moderate formula accuracy 60.27\% and template accuracy 61.52\% with BLEU scores (0.2824 on STL-DivEn, 0.5872 on DeepSTL). This indicates a balanced reward signal that moderately improves structural correctness and generation quality.

% Reward metric $m_t$ shows also shows moderate formula accuracy (61.72\% on STL-DivEn, 51.82\% on DeepSTL) and achieves BLEU score of 0.2652 on STL-DivEn and 0.5925 on DeepSTL, suggesting its effectiveness.

% Reward metric $m_l$ consistently underperforms across all metrics and datasets but still higher than the LLaMA3-8B (Fine-tuned), which is consistent with the results in Section Ablation Study.

% Reward metric $m_s$ performs best among single models in template accuracy and formula accuracy (62.82\% and 53.73\% respectively), and achieves strong BLEU scores. This shows that $m_s$ effectively promotes accuracy of generation.

\subsection{E.3 Accuracy of Reward Models}

\begin{table}[!t]
\centering
\renewcommand{\arraystretch}{1.3}
\resizebox{\linewidth}{!}{
\begin{tabular}{llccc}
\toprule
\textbf{Reward Model} & \textbf{Order} & \textbf{STL-DivEn}  & \textbf{DeepSTL}  & \textbf{Avg.}  \\
\midrule
\multirow{3}{*}{$r_a$ (AP Align)} 
    & Forward & \textbf{80.1} & \textbf{78.4} & \textbf{79.3} \\
    & Reverse & 74.6 & 73.1 & 73.9 \\
    & Shuffle & 76.9 & 75.5 & 76.2 \\
\midrule
\multirow{3}{*}{$r_t$ (NL Sim.)} 
    & Forward & \textbf{82.3} & \textbf{80.2} & \textbf{81.3} \\
    & Reverse & 76.2 & 74.4 & 75.3 \\
    & Shuffle & 78.5 & 76.7 & 77.6 \\
\midrule
\multirow{3}{*}{$r_l$ (Length)} 
    & Forward & \textbf{81.7} & \textbf{79.1} & \textbf{80.4} \\
    & Reverse & 74.1 & 72.5 & 73.3 \\
    & Shuffle & 77.2 & 74.9 & 76.1 \\
\midrule
\multirow{3}{*}{$r_s$ (STL Sim.)} 
    & Forward & \textbf{84.0} & \textbf{82.1} & \textbf{83.1} \\
    & Reverse & 77.6 & 75.9 & 76.8 \\
    & Shuffle & 79.3 & 77.4 & 78.4 \\
\bottomrule
\end{tabular}
}
\caption{Accuracy (\%) of reward models under different curriculum learning orderings on STL-DivEn and DeepSTL datasets.}
\label{tab:curriculum-accuracy}
\end{table}

We analyze the impact of curriculum learning on training reward models. As shown in Table~\ref{tab:curriculum-accuracy}, we evaluate the accuracy of four reward models: \( r_a \), \( r_t \), \( r_l \), and \( r_s \) in distinguishing between ``chosen" and ``rejected" outputs across two datasets, and compare three curriculum ordering strategies, which are the easy-to-hard (Forward), hard-to-easy (Reverse), and random shuffle (Shuffle).

The results demonstrate that the forward ordering achieves the highest accuracy. For instance, \( r_a \) achieves 80.1\% accuracy on the STL-DivEn dataset, outperforming  Reverse (74.6\%) and Shuffle (76.9\%).
% .whereas the reverse and shuffle orderings obtain lower accuracies of 74.6\% and 76.9\%, respectively. 
This trend is observed consistently across all reward models and both datasets, suggesting that progressively increasing task difficulty facilitates more effective stepwise learning and better capture of reward characteristics.

\subsection{F LLM Prompts for STL Transformation}

\begin{figure}
    \centering
    \includegraphics[width=1\linewidth]{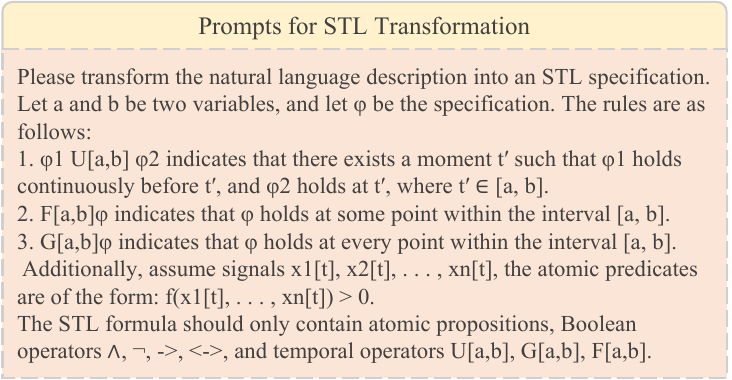}
    \caption{The prompts for STL Transformation.}
    \label{fig:prompt design}
\end{figure}

We provide the prompt to LLMs to guide the transformation of natural language into STL, as shown in Figure~\ref{fig:prompt design}.

% \begin{table}[!t]
% \centering
% \begin{tabular}{p{0.95\linewidth}}
% \toprule
% Please transform the natural language description into an STL specification. Let \(a\) and \(b\) be two variables, and let \(\varphi\) be the specification. The rules are as follows:

% 1. \(\varphi_1 \,\mathcal{U}_{[a,b]}\, \varphi_2\) indicates that there exists a moment \(t'\) such that \(\varphi_1\) holds continuously before \(t'\), and \(\varphi_2\) holds at \(t'\), where \(t' \in [a,b]\).

% 2. $\F_{[a,b]} \varphi$ indicates that \(\varphi\) holds at some point within the interval \([a,b]\).

% 3. $\G_{[a,b]} \varphi$ indicates that \(\varphi\) holds at every point within the interval \([a,b]\).

% Additionally, assume signals \(x_1[t], x_2[t], \ldots, x_n[t]\), the atomic predicates are of the form:
% \[
% f(x_1[t], \ldots, x_n[t]) > 0.
% \]
% The STL formula should only contain atomic propositions, Boolean operators \(\wedge, \neg, \to, \leftrightarrow\), and temporal operators \(\mathcal{U}_{[a,b]}\), $\G_{[a,b]}$, $\F_{[a,b]}$.
% \\
% \bottomrule
% \end{tabular}
% \caption{Prompts for LLMs to guide NL-to-STL transformation}
% \label{tab:prompt_design}
% \end{table}

\end{document}